\documentclass{pasj00}

\begin{document}
\SetRunningHead{Kataoka et al.}{Suzaku Observations of 3C~120}
\Received{2006/11/28}
\Accepted{2006/12/28}

\title{Probing the Disk-jet Connection of the Radio Galaxy \\
3C~120 Observed with Suzaku}

\author{
   Jun \textsc{Kataoka}\altaffilmark{1}
   James N. \textsc{Reeves}\altaffilmark{2}
   Kazushi \textsc{Iwasawa}\altaffilmark{3}
   Alex G. \textsc{Markowitz}\altaffilmark{2}\\
   Richard F. \textsc{Mushotzky}\altaffilmark{2}
   Makoto \textsc{Arimoto}\altaffilmark{1}  
   Tadayuki \textsc{Takahashi}\altaffilmark{4}\\ 
   Yoshihiro \textsc{Tsubuku}\altaffilmark{1}  
   Masayoshi \textsc{Ushio}\altaffilmark{4}
   Shin \textsc{Watanabe}\altaffilmark{4}\\
   Luigi C. \textsc{Gallo}\altaffilmark{3,4}
   Greg M. \textsc{Madejski}\altaffilmark{5}
   Yuichi \textsc{Terashima}\altaffilmark{6}\\
   Naoki  \textsc{Isobe}\altaffilmark{7}
   Makoto S. \textsc{Tashiro}\altaffilmark{8}
   and
   Takayoshi \textsc{Kohmura}\altaffilmark{9}
}
 \altaffiltext{1}{Department of Physics, Tokyo Institute of Technology,
   Meguro, Tokyo, Japan, 152-8551}
 \email{kataoka@hp.phys.titech.ac.jp}
 \altaffiltext{2}{NASA Goddard Space Flight Center, Greenbelt, MD 20771,
   USA}
 \altaffiltext{3}{Max Planck Institut f\"{u}r extraterrestrische
   Physik(MPE), Garching, Germany}
 \altaffiltext{4}{Institute of Space and Astronautical Science, JAXA,
   Sagamihara, Kanagawa, Japan, 229-8510}
 \altaffiltext{5}{Stanford Linear Accelerator Center, Stanford, CA,
   943099-4349, USA}
 \altaffiltext{6}{Department of Physics, Ehime University, 
   Matsuyama, Ehime, 790-8577, Japan}
 \altaffiltext{7}{Cosmic Radiation Laboratory, Institute of Physical and
   Chemical Research, Saitama, Japan}
 \altaffiltext{8}{Department of Physics, Saitama University, Saitama, Japan, 338-8570}
 \altaffiltext{9}{Physics Department, Kogakuin University, 
  Hachioji, Tokyo, Japan, 192-0015}
\KeyWords{galaxies: individual (3C~120); galaxies: active;
galaxies: Seyfert; X-rays: galaxies} 

\maketitle

\begin{abstract}

Broad line radio galaxies (BLRGs) are a rare type of radio-loud AGN, in 
which the broad optical permitted emission lines have been detected in 
addition to the extended jet
emission. Here we report on deep (40ksec $\times$4) observations 
of the bright BLRG 3C~120 using Suzaku. The observations were 
spaced a week apart, and sample a range of continuum fluxes.
An excellent broadband spectrum was obtained over  
two decades of frequency (0.6 to 50 keV) within each 40 ksec 
exposure. We clearly resolved the iron K emission line complex, finding that
it consists of a narrow  K$_{\alpha}$ core ($\sigma$ $\simeq$ 110 eV or an EW of 60 eV), 
a 6.9 keV line, and an underlying broad iron line. Our confirmation
of the broad line contrasts with the XMM-Newton observation in 2003,
where the broad line was not required.
The most natural interpretation of the broad line is  
iron K line emission from a face-on accretion disk which is truncated at 
$\sim$ 10 $r_g$. Above 10 keV, a relatively weak Compton hump was detected 
(reflection fraction of $R$ $\simeq$ 0.6), superposed on the primary X-ray 
continuum of $\Gamma$ $\simeq$ 1.75. Thanks to the good photon statistics
and low background of the Suzaku data, we clearly confirm the spectral 
evolution of 3C~120, whereby the variability amplitude decreases with 
increasing energy.  More strikingly, we discovered that the variability 
is caused by a steep power-law component of $\Gamma$ $\simeq$
2.7, possibly related to the non-thermal jet emission. We discuss 
our findings in the context of similarities and differences between 
radio-loud/quiet objects. 

\end{abstract}

\section{Introduction}

One of the most important issues in the study of active galactic 
nuclei (AGN) is why well-collimated, powerful, relativistic radio 
jets exist only in 10 $\%$ of the AGN class, i.e., in $so$-$called$ 
radio-loud objects (e.g., Urry \& Padovani 1995).\footnote{More accurately, 
a well-known dichotomy of the radio-loud and radio-quiet 
objects is based on the $old$ VLA study of PG quasars 
(Kellermann et al. 1989), and recent deep VLA FIRST survey shows 
no such sign of bimodality in radio loudness (White et al. 2000).} 
All AGNs are thought to be powered by accretion of matter onto a 
supermassive black hole, presumably via an equatorial 
accretion disk. Recent VLBI observations of a nearby 
active galaxy M87 confirmed 
that the jet is $already$ launched within $\sim$60 $r_g$ 
(where $r_g$ = $GM$/c$^2$ is the gravitational radius), 
with a strong collimation occurring within $\sim$200 $r_g$ 
of the central black hole (Junor et al.\ 1999). 
These results are consistent with the hypothesis that jets are formed 
by an accretion disk, which is threaded by a magnetic field. 
Therefore the observational properties of the accretion disk 
and corona are essential ingredients to jet formation 
(e.g., Livio 1999 and references therein).

The profile of the iron K$_{\alpha}$ (6.4 keV) line can be used to 
probe the structure of the accretion disk, because it is thought to result 
from fluorescence of the dense gas in the geometrically thin and
optically thick regions of the inner accretion disk 
($\sim$ 10 $r_{\rm g}$). The most famous example is the spectrum of 
the Seyfert 1 (Sy-1) galaxy MCG--6-30-15, which shows a relativistically 
broadened Fe K$_{\alpha}$ emission line, first detected by ASCA 
(Tanaka et al.\ 1995; Iwasawa et al.\ 1996; 1999).  This finding 
was confirmed by  Chandra and XMM-Newton (e.g., Wilms et al.\ 2001; Fabian et
al. 2002), and most recently by Suzaku (Minuitti et al.\ 2006).  
Similar broad relativistic iron line profiles have
been detected in several other type-1 AGN (e.g., 
Nandra et al.\ 1999, Iwasawa et al.\ 2004, Turner et al.\ 2005 for 
NGC 3516; Page et al.\ 2001 for Mrk 766), although they are perhaps 
somewhat less common than anticipated from the ASCA 
era (e.g., Nandra et al.\ 1997). In contrast, the presence of a narrower 
6.4 keV line from more distant matter (e.g., from the outer disk, 
broad/narrow line regions, or torus) is common in many type-1 AGNs 
(e.g., Yaqoob \& Padmanabhan 2004). In this context, studies of the  
iron line profile in radio-loud AGN provides important clues to the 
disk-jet connection, particularly by comparison with Sy-1s.
Moreover, in  a standard picture of two-phase disk-corona 
model proposed by Haardt \& Maraschi  (1991), the corona 
temperature is related to the high energy cutoff 
typically observed  in the hard X-ray band 
(see also Poutanen \& Svennson 1996). Thus observations of 
the Compton reflection hump and its cutoff above 10 keV  
can be used to determine the geometry and temperature of 
the disk-corona system postulated to exist in both 
types of AGNs (e.g., Wo\'{z}niak et al.\ 1998).

3C~120 ($z$ = 0.033) is the brightest broad line radio galaxy (BLRG), 
exhibiting characteristics intermediate between those 
of FR-I radio galaxies and BL Lacs. It has a one-sided superluminal 
jet on 100 kpc scales (Walker et al.\ 1987; Harris et al.\ 2004),  
and superluminal motion (with an apparent velocity $\beta_{\rm
app}$ = 8.1) has been observed for the jet component (Zensus 1989).
This provides an upper limit to the inclination angle of the jet to 
the line of sight of 14 deg (Eracleous \& Halpern 1988). Interestingly, 
the optical spectrum of 3C~120 is $not$ LINER-like, as is often seen 
in FR-I radio galaxies (e.g., Baum et al.\ 1995), but rather typical of 
Sy-1s. It resides in an optically peculiar galaxy that shows only 
some indication of spiral structure (Moles et al.\ 1988). From 
reverberation mapping, the central black hole mass is remarkably well 
constrained: $M$ = 5.5$^{+3.1}_{-2.3}$ $\times$10$^{7}$ $M_{\odot}$ 
(Peterson et al.\ 2004; see also Wandel, Peterson \& Malkan 1999). 
Very recently, Marscher et al.\ (2002) 
found that dips in the X-ray emission of 3C~120 are followed by 
ejections of bright superluminal knots in the radio jet, which 
clearly indicates an important connection between the jet and the 
accretion disk. 

In X-rays, 3C~120 has been known to be a bright ($\sim$ 5$\times$10$^{-11}$  
erg cm$^{-2}$ s$^{-1}$ at 2$-$10 keV), variable source with a canonical 
power-law spectral shape that softened as the source brightened (e.g., 
Maraschi et al.\ 1991 and reference therein). A broad iron K$_{\alpha}$
line was first detected by ASCA in 1994, with its width 
$\sigma$ = 0.8 keV and EW (equivalent width) $\sim$400 eV 
(Grandi et al.\ 1997; Reynolds 1997; Sambruna et al.\ 1999). 
The follow-up observations by RXTE (Eracleous, 
Sambruna \& Mushotzky 2000; Gliozzi, Sambruna \& Eracleous 2003)  and 
BeppoSAX (Zdziarski \& Grandi 2001) also detected  a broad iron line,
but with a much smaller EW of $\sim$100 eV. Furthermore, these observations have 
confirmed a presence of a weak Compton hump in 3C~120, with a reflection 
normalization of $\Omega$/2$\pi$ $\sim$ 0.4.  
It was argued that both the weak line 
and relative weakness of the Compton hump is suggestive of an optically 
thick accretion disk which transitions to a hot, optically thin flow 
(Eracleous, Sambruna \& Mushotzky 2000; Zdiarski \& Grandi 2001).
Without a doubt, 3C~120 is a key source to construct a unified view of 
radio-loud and radio-quiet objects.

Most recently, 3C~120 was observed for nearly a full orbit (130 ksec) 
with XMM-Newton on 26$-$27 August 2003 (Ballantyne et al.\ 2004, 
Ogle et al.\ 2005). This clearly confirmed the presence of the neutral Fe line 
emission (57$\pm$7 eV in EW), which was slightly broadened with a FWHM of $\sigma$ 
= 9000$\pm$3000 km s$^{-1}$.  Both of these papers  argued that the line 
profile is rather symmetric and no evidence was found for  
relativistic broadening, or alternatively the line arises from an accretion 
disk radius of $\ge$75 $r_g$ at an inclination angle of $\sim$10 deg
(where relativistic gravitational effects are almost negligible). 
They also confirmed a weaker emission 
line at 6.9 keV with EW = 20$\pm$7 eV (initially suggested by Chandra HETGS;
Yaqoob \& Padmanabhan 2004), which can be interpreted as a 
blend of Fe K$_{\beta}$ and H-like (or He-like) iron lines. 
Despite significant progress made by XMM-Newton, most of the results are not
conclusive due to XMM-Newton's limited energy range (no coverage above 
10 keV), and the relatively high background of XMM-Newton above 5 keV. 
At present, various models can fit these iron lines equally well,
though they assume quite different geometries and/or 
ionization states (Ballantyne et al.\ 2004, Ogle et al.\ 2005). 
It has been argued that such degeneracies may
be resolved by Suzaku, due to its unprecedented sensitivity between 0.3 and 
$\sim$300 keV.  

In this paper, we present a detailed analysis of 160 ksec worth of 
data on 3C~120, observed with Suzaku in February and March 2006 as a part of 
the SWG (science working group) program. As our aim was to 
monitor the source in various states of source activity, we divided 
this total exposure into four exposures of 40 ksec each, with one pointing 
per week (to be roughly equal to the variability timescale of this source;
e.g., Marshall et al.\ 2003; Gliozzi et al.\ 2003). 
Thanks to the excellent energy resolution and sensitivity of 
the XIS and HXD/PIN onboard Suzaku, we successfully obtained one of 
the highest quality data on this radio galaxy ever reported, between 0.6 and
50 keV. The paper is organized as follows. 
The observation and analysis methods are
described in $\S$2. We present an overview of 3C~120's variability during 
the Suzaku observations and temporal studies of light curves in $\S$3. 
Detailed spectral studies are presented in $\S$4; in particular we focus on 
(1) multiband spectral features,  (2) the nature of the iron K line complex,
and (3) the difference spectrum between high and low states. Based on these 
new findings,  we discuss the nature of the variability and spectral evolution
of 3C~120 in $\S$5. Finally, we present our conclusions in
$\S$6. Uncertainties of background models on the HXD/PIN light curve 
will be discussed further in detail in the appendix.

\section{Observation and Data Reduction}

\begin{table*}
  \caption{Suzaku observation log of 3C~120.}\label{tab:first}
  \begin{center}
    \begin{tabular}{ccccc}  
    \hline
    obs\_ID & start (UT) & stop (UT) & XIS/HXD exp. (ksec) \\
    \hline
    3C~120 $\#$1  & 2006 Feb 09  03:20 & 2006 Feb 10 05:50 & 33.3/29.3 \\  
    3C~120 $\#$2  & 2006 Feb 16  13:08 & 2006 Feb 17 14:06 & 36.4/32.6 \\  
    3C~120 $\#$3  & 2006 Feb 23  20:02 & 2006 Feb 24 18:00 & 37.0/36.2 \\  
    3C~120 $\#$4  & 2006 Mar 02  22:29 & 2006 Mar 03 20:39 & 37.9/37.4 \\  
    \hline
    \end{tabular}
  \end{center}
\end{table*}

The broad line radio galaxy 3C~120 was observed with Suzaku 
(Mitsuda et al.\ 2006) four times in February and March 2006 with a 
total (requested) duration of 160 ksec.  
Table 1 summarizes the start time and the 
end time, and the exposures of each observation. 
Suzaku carries four sets of X-ray telescopes 
(Serlemitsos et al.\ 2006) each with a focal-plane X-ray CCD camera 
(XIS, X-ray Imaging Spectrometer; Koyama et al.\ 2006) that is sensitive 
in the energy range of 0.3$-$12 keV, together with a non-imaging 
Hard X-ray Detector (HXD, Takahashi et al.\ 2006; Kokubun et al.\ 2006), 
which covers the 10$-$600 keV energy band with Si PIN photo-diodes and GSO 
scintillation detectors. Three of the XIS (XIS 0, 2, 3) have 
front-illuminated (FI) CCDs, while the XIS 1 utilizes a back-illuminated (BI) 
CCD. The merit of the BI CCD is its improved sensitivity in the soft X-ray
energy band below 1 keV. In all four observations (3C~120 $\#1 - 4$), 
3C~120 was focused on the nominal center position of the XIS detector.

\subsection{XIS Data Reduction and Analysis }

For the XIS, we analyzed version 0.7 of the screened data (Fujimoto et al.\ 2006) 
provided by the Suzaku team. The screening of the version 0.7 data are 
based on the following criteria; (1) only ASCA-grade 0,2,3,4,6 events 
(Yamaguchi et al.\ 2006) 
are accumulated, while hot and flickerling pixels were removed from 
the XIS image using the \textsc{CLEANSIS}  script 
(2) the time interval after the passage of South Atlantic Anomaly 
(T\_SAA) is larger than 436 sec, (3) the object is at 
least 5 deg and 20 deg above the rim of the Earth (ELV) during night 
and day, respectively.  In addition, we also select the data with the cutoff 
rigidity (COR) larger than 6 GV. The XIS events were extracted 
from a circular region with a radius of 4.3' centered on the source
peak, whereas the background was accumulated in an annulus with 
its inner and outer radii of 4.9' and 6.6', respectively. 
For the  3C~120 $\#3$ observation (see Table 1), we discard the 
first 600 sec data after the maneuver  because the pointing 
fluctuated from its center (i.e., XIS nominal position) by $\sim$ 2'.

The XIS spectra were corrected for the hydrocarbon 
(C$_{24}$H$_{38}$O$_{4}$) contamination on the optical blocking 
filter, by including an extra absorption column due to Carbon and 
Oxygen in all the spectral fits. The column densities for each 
detector were calculated based on the date of the observation 
using an empirical relation released by the Suzaku instrumental 
team. We assumed different Carbon and Oxygen column densities 
($N_{\rm C}$ and $N_{\rm O}$) for different sensors and observations;  
$N_{\rm C}$ = (2.38$-$2.56)$\times$10$^{18}$, 
(3.20$-$3.50)$\times$10$^{18}$, (3.65$-$3.80)$\times$10$^{18}$, 
and (5.44$-$5.65)$\times$10$^{18}$ atoms cm$^{-2}$ for XIS 0, 1, 2, 3 
respectively, with the ratio of C/O column densities 
($N_{\rm C}$/$N_{\rm O}$) set to 6. This additional soft X-ray 
absorption due to the hydrocarbon contamination was included as a 
fixed spectral component using the \textsc{varabs} absorption model in 
all the spectral fits.

Although 3C~120 is known to have an large extended X-ray jet that 
is $\sim$20 arcsec apart from the nucleus (Harris et al.\ 2004), 
and could not be resolved with Suzaku, the X-ray contribution 
from this large-scale (10 kpc scale) jet is less 
than 1 $\%$ of the nucleus emission 
in the 0.5$-$10 keV bandpass ($F_{\rm jet}$ 
$\sim$ 2.3$\times$10$^{-13}$ erg cm$^{-2}$ s$^{-1}$). 
We therefore use the latest version of both the response matrix and the point 
spread function (PSF)  released by the Suzaku team for the point-like
sources, ae\_xi[0,1,2,3]\_20060213.rmf and 
ae\_xi[0,2,3]\_xisnom6\_20060415.arf.  Since the nuclear emission 
is very bright from 0.5$-$10 keV, we binned the XIS spectra to a 
minimum of 200 counts per bin to enable the use of the $\chi^2$ 
minimization statistic.

\begin{figure}
  \begin{center}
    \FigureFile(85mm,85mm){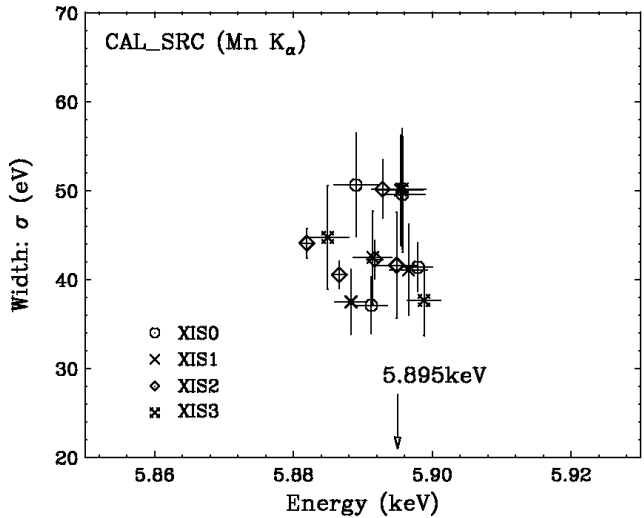}
  \end{center}
  \caption{Energy scale and residual width of the $^{55}$Fe calibration source
 on each of the XIS sensors, during each of the four observations.}\label{fig:Fe-cal}
\end{figure}

Another note on the XIS data analysis is the accuracy of the energy scale 
(reported to be less than  0.2 $\%$; Koyama et al.\ 2006) and energy 
resolution. These calibrations are very important because one of 
our ultimate goals is to obtain the precise measurement of the iron line 
complex, as we will see in $\S$ 4.3. 
For this purpose, the $^{55}$Fe calibration source 
located on the corners of the XIS chips were used as an accurate 
calibrator of the instrumental response during 3C~120 observations.
\footnote{Note the $^{55}$Fe calibration source is on the 
corners of the XIS chips, while 3C120 is focused on the center of the 
CCD. From detailed analysis of galaxy clusters and Sgr C, the XIS
team confirmed this position difference makes only negligible effects in 
both the line width and spatial gain variations.}
The $^{55}$Fe source produces a characteristic X-ray line from 
Mn K$_{\alpha}$ at 5.895 keV (a combination of  K$_{\alpha 1}$ 
and K$_{\alpha 2}$ at 5.899 keV and 5.888 keV respectively with a 
branching ratio of 2:1). Figure \ref{fig:Fe-cal} 
shows the results of Gaussian fitting 
for each of the XIS chips during 3C~120 observation $\#1 - 4$. 

From measuring the 5.895 keV lines in the calibration source, 
we find that the line energy is shifted slightly redwards by 
5.5$\pm$0.6 eV, while there is some residual width (after the 
instrumental response function has been accounted for) in the
calibration lines of $\sigma_{\rm cal}$ = 45$\pm$1 eV. 
A further check of the line width was made by using  
Sgr C (molecular cloud) data which was 
taken between 3C~120 $\#$2 and $\#$3. As reported in Ueno et
al.\ (2006), the 1 $\sigma$ width of the 6.4 keV line in the Sgr C 
observation was 39$\pm$5 eV, which is consistent with the 3C~120 calibration lines
within the statistical error. Since the intrinsic line width is 
expected to be negligible for this molecular cloud, it must be 
mostly instrumental.  Therefore 
in the following, the intrinsic width of the iron K$_{\alpha}$ line 
is simply evaluated as $\sigma_{\rm int}$ = $\sqrt{\sigma_{\rm obs}^2 - 
\sigma_{\rm cal}^2}$, where $\sigma_{\rm obs}$ 
is the measured width of the iron line discussed below.

\subsection{HXD Data Reduction and Analysis}

The source spectrum and the light curves were extracted from the 
cleaned HXD/PIN event files (version 1.2) provided by the Suzaku team. 
The HXD/PIN data are processed with the screening criteria basically 
the same as those for the XIS, except that ELV $\ge$ 5 deg through 
night and day, 
T\_SAA $\ge$ 500 sec, and COR $\ge$ 8 GV. The HXD/PIN instrumental 
background spectra were generated from a time dependent 
model provided by the HXD instrument team for each observation. 
The model utilized the count rate of upper discriminators and COR values 
as the measure of cosmic-ray flux that pass through the Si PIN 
diode; background spectra based on a database of non 
X-ray background observations made by the PIN diode to date are provided   
(See Kokubun et al.\ 2006; Fukazawa et al.\ 2006 for more details).
At the time of writing, two different background models, A and B, are 
under investigation, but the difference
between the two models is rather small for our 3C~120 HXD/PIN data analysis. 
According to careful examinations using simulations on various
observational datasets, the uncertainty of the background models for 
the PIN detector is expected to be $\sim$ 5$\%$ for both model-A and B 
(see Appendix). 

Both the source and background spectra were made with identical GTIs 
(good time intervals) and the exposure was corrected for 
detector deadtime of $\sim$ 5$\%$ (ranges in 3.5$-$8.5 $\%$ for 
3C~120 observations; see Appendix). 
To minimize uncertainties due to limited photon statistics on 
the background models, 
background spectra were generated 
with 10 times the actual background count rate but increasing the
effective exposure time of background by a factor of 10. The 
HXD/PIN response file dated 2006-08-14 for the XIS nominal position 
(ae\_hxd\_pinxinom\_20060814.rsp) was used for these spectral fits. 
Full details of the HXD instrument and performance are given 
in Takahashi et al.\ (2006) and Kokubun et al.\ (2006).

\begin{figure}
  \begin{center}
    \FigureFile(85mm,85mm){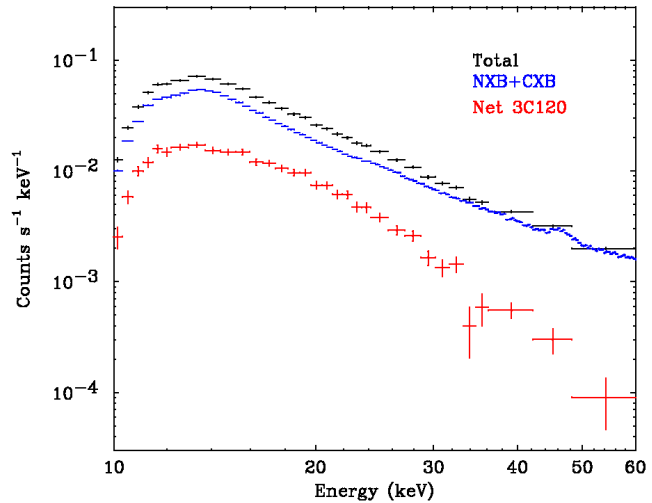}
  \end{center}
  \caption{The combined HXD/PIN spectra for the Suzaku observation of 3C~120 
$\#1 - 4$ over the whole HXD/PIN energy bandpass (10$-$60 keV). 
Black shows the total spectrum (source plus background),
 blue shows sum of non-X-ray background and the CXB, and red for the net 
source spectrum. Note the source is detected in each data bin at more
 than 10 $\%$ of background level over the 12$-$50 keV range used in this 
paper.
)}\label{fig:pinspec}
\end{figure}

The time averaged (obs $\#1 - 4$ combined) HXD/PIN spectrum thus 
obtained is shown in Figure \ref{fig:pinspec}, plotted over the energy
range of 10 to 60 keV. HXD/PIN data below 12 keV have been ignored to 
avoid noise contamination near the lower threshold of the PIN diode. 
Also the data above 50 keV are discarded, as a detailed study of noise 
and background systematics is on-going above this energy. 
Figure \ref{fig:pinspec} shows the total 
(3C~120 + observed background) spectrum, where the background 
includes both the instrumental (non X-ray) background model-A and the 
contribution from the cosmic X-ray background (CXB) (Gruber et
al. 1999).  Here the form of the CXB was taken as 
9.0$\times$10$^{-9}$(E/3keV)$^{-0.29}$ exp($-$E/40keV) 
erg cm$^{-2}$ s$^{-1}$ sr$^{-1}$ keV$^{-1}$ and the 
observed spectrum was simulated assuming the PIN detector response 
to isotropic diffuse emission. When normalized to the field of 
view of the HXD/PIN instrument the effective flux of the CXB component 
is expected to be 9.0$\times$10$^{-12}$ erg cm$^{-2}$ s$^{-1}$ 
in the 12$-$50 keV band, which is about $\sim$14 $\%$ of the 3C~120 
flux in the same energy bandpass. 

3C~120 is known to have a 50$-$100 keV flux of $\sim$ 5$\times$10$^{-11}$ 
erg cm$^2$ s$^{-1}$ (e.g., Wo\'{z}niak et al.\ 1998; Zdziarski \& Grandi
2001) and therefore ultimately can be detected 
by the HXD/GSO detector. However, this is only a few percent 
of the GSO detector background and the study at this level of 
background systematics is still on-going by the HXD instrumental team. 
Therefore in this paper, 
we do not include the GSO data in the subsequent spectral fits.
Above 50 keV, a reliable detection of 3C~120 cannot be made at the 
present time by the GSO, but results using new 
response matrices and revised background models will be discussed 
elsewhere in the near future.

\section{Temporal Studies}
\subsection{Overview of Variability}

\begin{figure}
  \begin{center}
    \FigureFile(85mm,85mm){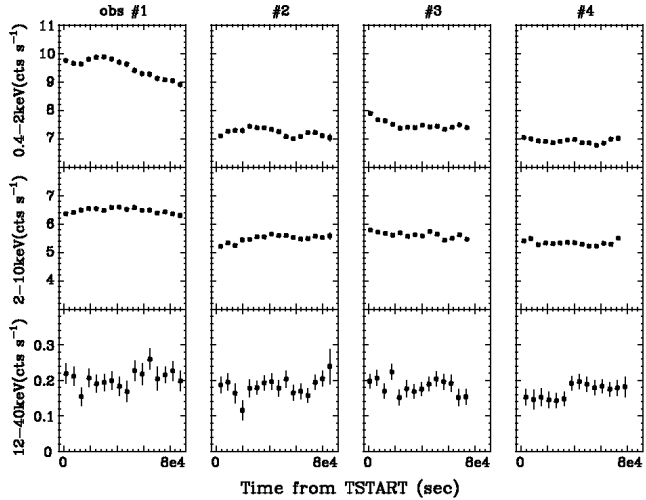}
  \end{center}
  \caption{The overall variability of 3C~120 during February and March  
observations by Suzaku.  $upper$: 0.4$-$2 keV XIS (XIS 0$-$3
summed), $middle$: 2$-$10 keV XIS (XIS 0$-$3
summed), and $lower$: 12$-$40 keV HXD/PIN. The backgrounds 
are subtracted.}\label{fig:lc}
\end{figure}

Figure \ref{fig:lc} shows an overview of the count rate variations of 3C~120 
during the February  and March observations ($\#1 - 4$). The light curves of the
4 XISs and PIN detectors are shown separately in different 
energy bands; 0.4$-$2 keV ($upper$; XIS), 2$-$10 keV ($middle$; XIS) and 
12$-$40 keV ($lower$; HXD/PIN). 
The net source count rates, averaged over four observations,  
measured from 0.4$-$10 keV were 2.965$\pm$0.005 cts s$^{-1}$, 
3.703$\pm$0.005 cts s$^{-1}$, 2.972$\pm$0.005 cts $^{-1}$, and
2.799$\pm$0.005 cts $^{-1}$,  
respectively for the XIS 0, 1, 2, 3. The background is typically 3 $\%$ of 
the source counts for the FI-XIS (XIS 0,2,3) and 7 $\%$ for the BI-XIS
(XIS 1). For the PIN detector, the net average 
source count rate in the 12$-$40 keV band was 0.150$\pm$0.002 cts $^{-1}$, 
compared to 
the PIN background (non X-ray background + CXB) rate of 0.449 cts $^{-1}$. 

Figure \ref{fig:lc} clearly indicates that 3C~120 was 
in a relatively high state during the 1st 40 ksec observation 
($\#1$; the summed count rates of 4 XISs detectors was 
15.94$\pm$0.01 counts s$^{-1}$), 
then its count rate dropped by $\sim$20$\%$  in the 2nd observation, 
and finally reached a minimum in the 4th observation 
($\#4$; 12.02$\pm$0.01 counts s$^{-1}$). 
Here the effect of time dependent degradation of the XIS efficiency 
due to C and O contamination (Koyama et al.\ 2006; see also $\S$2.1) 
is corrected between 3C~120$\#$1 and $\#$4, though this makes  only 
negligible effect even below 2 keV within a month scale.
Interestingly, a large time variation of the 0.4$-$2 keV count rate in the 1st 
observation was not clearly detected above 2 keV, suggesting a relative 
lack of variability at higher
energies, as has already been noticed for this particular source 
by various authors 
(e.g., Maraschi et al.\ 1991 and Zdziarski \& Grandi 
2001). The count rate variations of the HXD/PIN detector is less clear due to 
limited photon statistics, but some level of variability exists, as  
we will see in the next section.

\begin{figure}
  \begin{center}
    \FigureFile(75mm,75mm){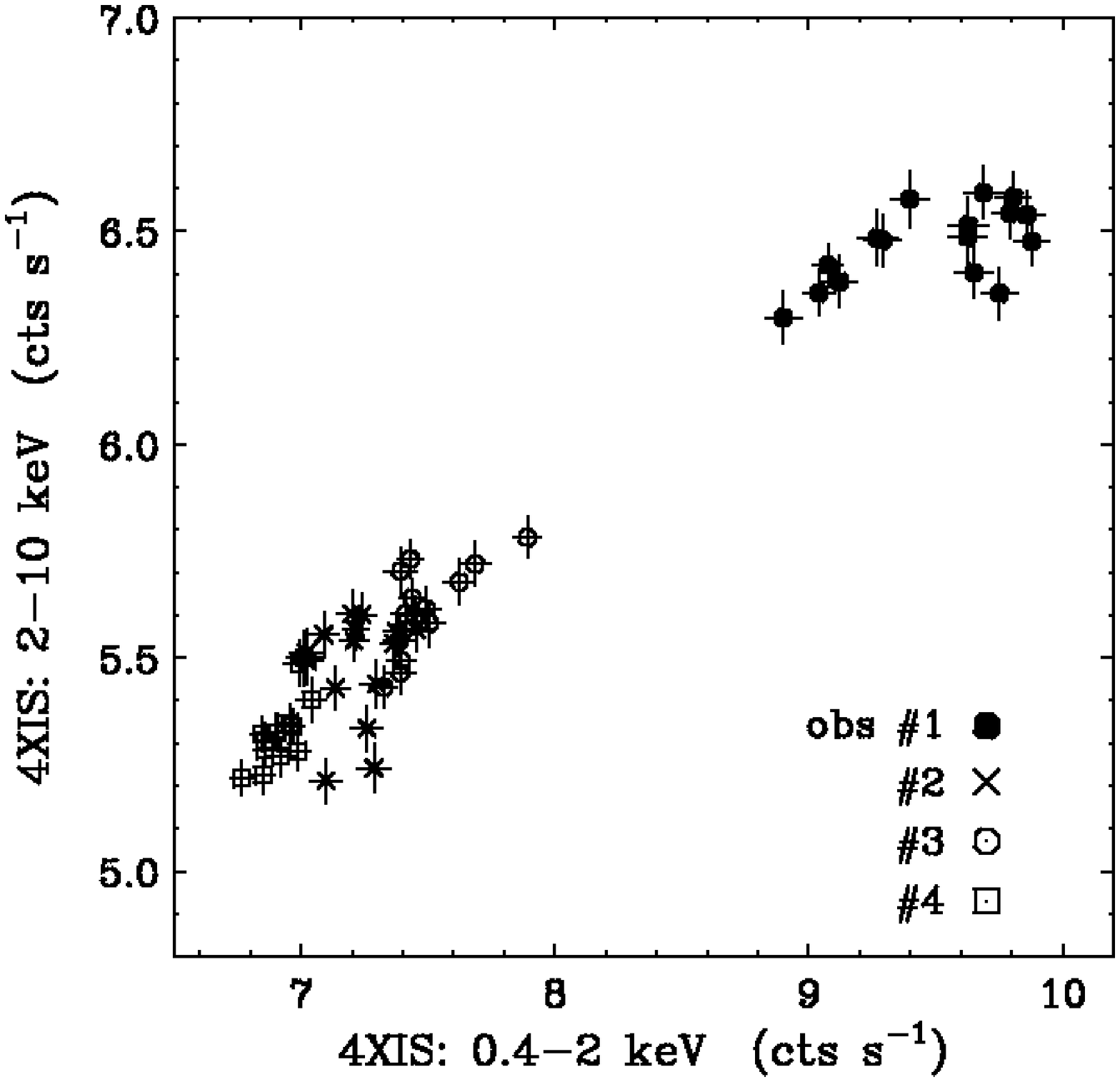}
    \FigureFile(75mm,75mm){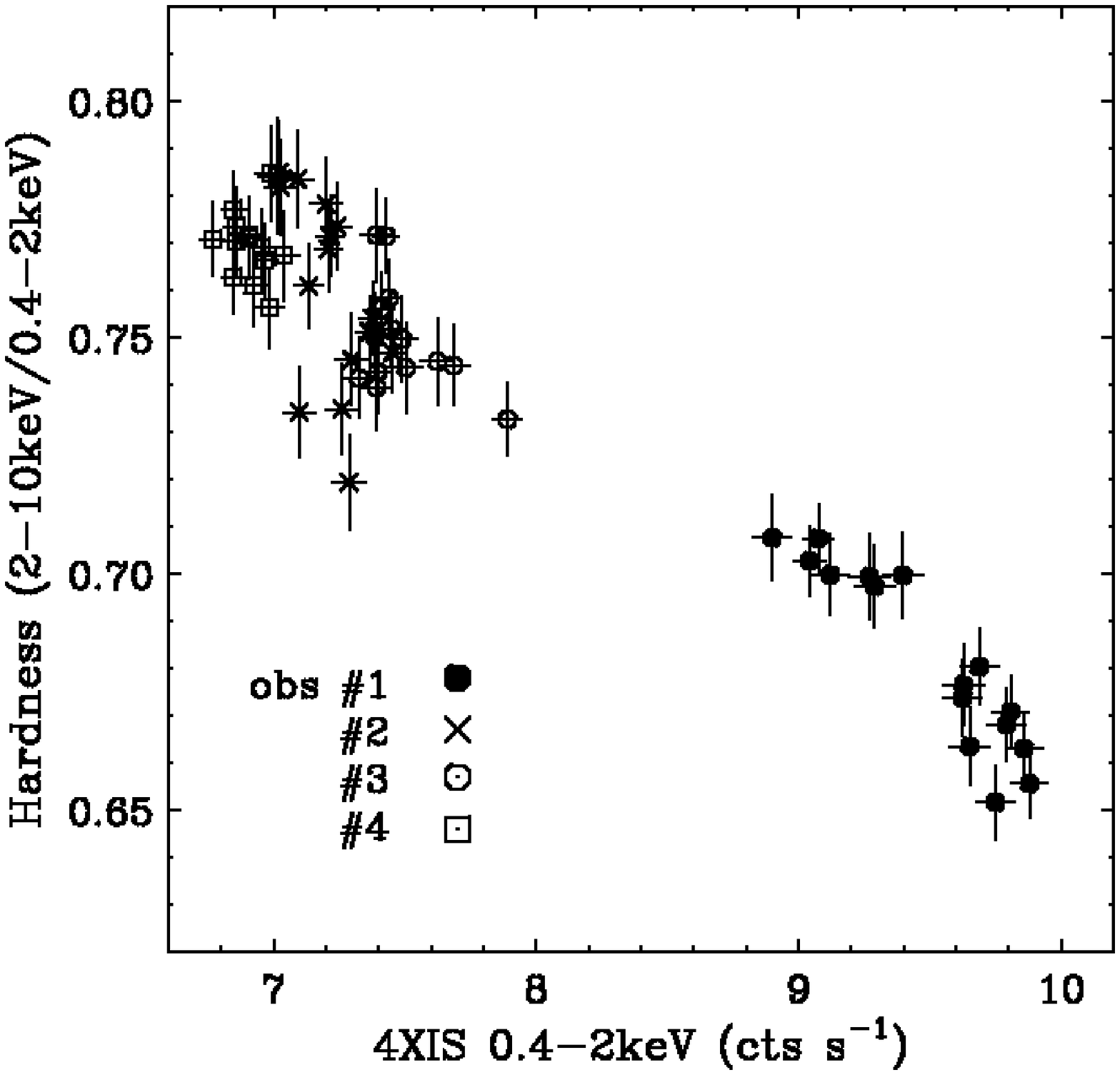}
  \end{center}
  \caption{(a) $upper$: Correlation of XIS count rates between 0.4$-2$
 keV  and 2$-$10 keV. (b) $lower$: Changes of hardness ratio between 0.4$-2$
 keV  and 2$-$10 keV. The hardness is defined  as 2$-$10 keV count rates 
divided by 0.4$-$2 keV count rates. }\label{fig:hd-xis}
\end{figure}

\subsection{Hardness Ratio}

\begin{figure}
  \begin{center}
    \FigureFile(75mm,75mm){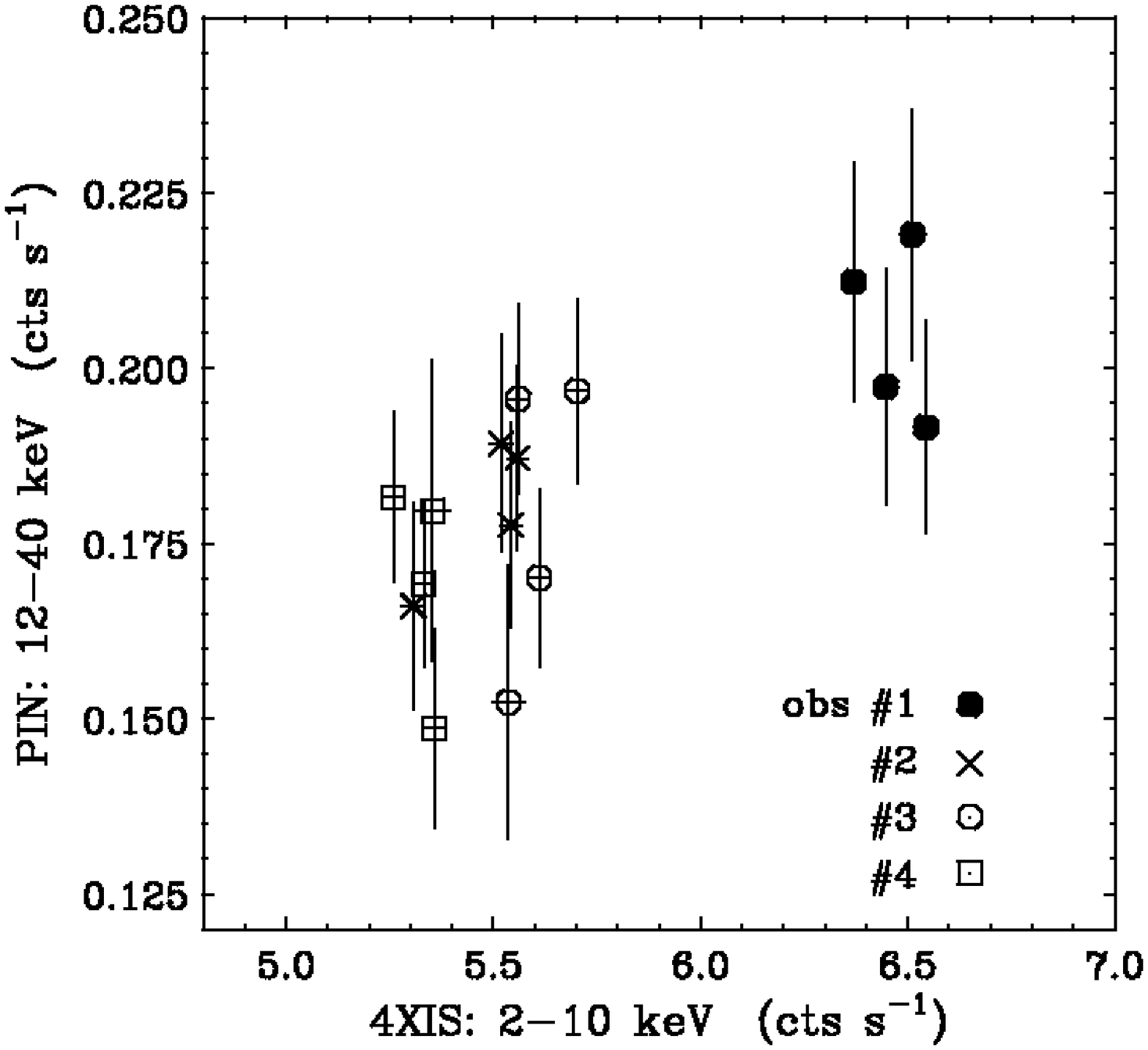}
    \FigureFile(75mm,75mm){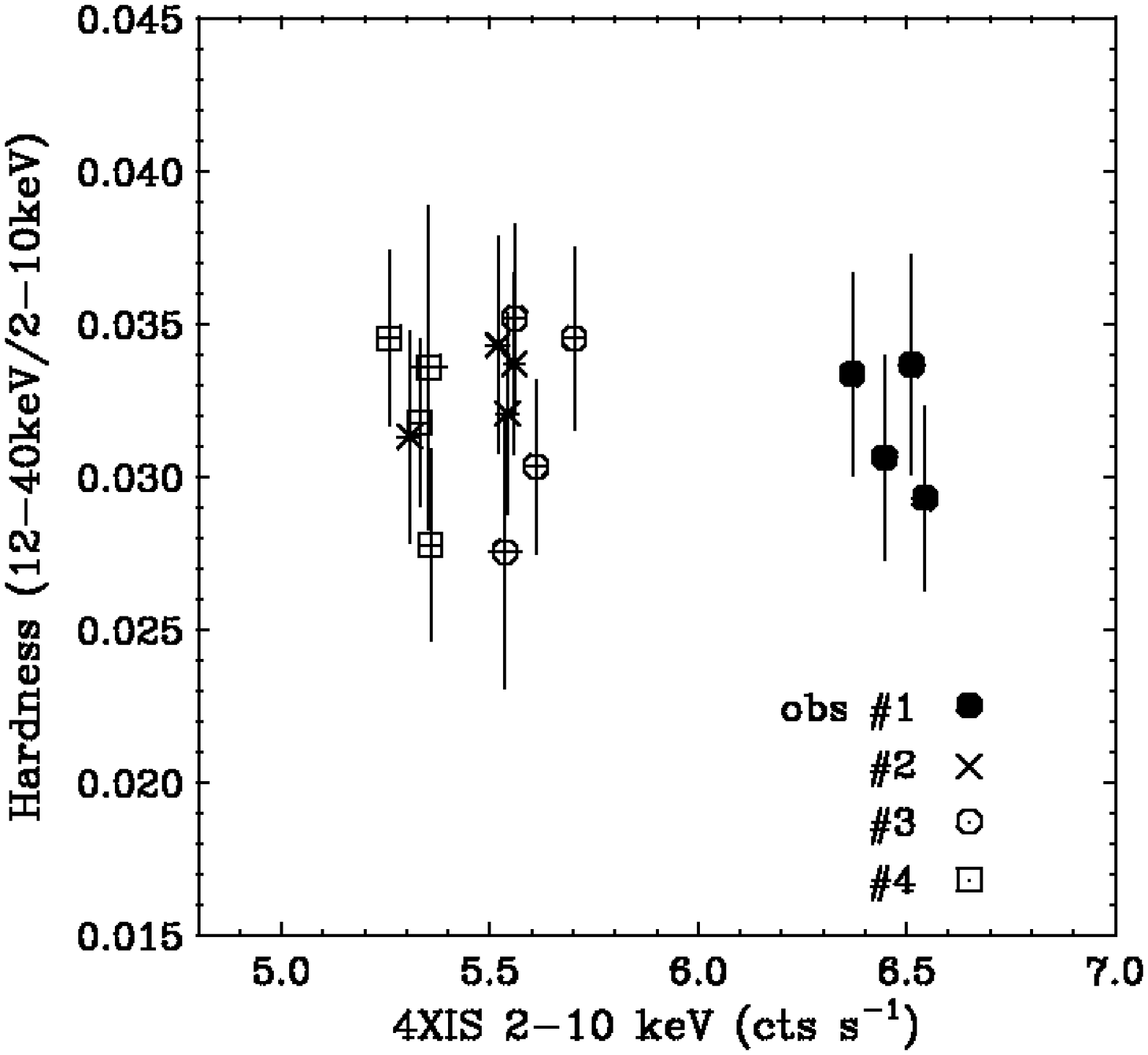}
  \end{center}
  \caption{(a) $upper$: Correlation of count rates between 2$-$10 
 keV (XISs) and 12$-$40 keV (HXD/PIN). (b) $lower$: Changes of hardness 
ratio between 2$-10$ keV  and 12$-$40 keV. The hardness is defined 
as the 12$-$40 keV count rate divided by the 2$-$10 keV count rate. }
\label{fig:hd-pin}
\end{figure}

Figure \ref{fig:hd-xis} 
(a) compares the XIS count rate correlation between the soft 
X-ray (0.4$-$2.0 keV) and hard X-ray energy bands (2$-$10 keV). 
Generally the correlation is tight,  such that hard X-ray flux
increases when the soft X-ray flux increases. However, some scatter in 
the correlation suggests some variation of the spectrum, even if 
the source is in similar flux states. Moreover, the amplitude 
of flux variations 
is approximately 40 $\%$ in 0.4$-$2 keV, but only $\sim$ 25 $\%$ in 2$-$10
keV. Figure \ref{fig:hd-xis} (b) 
shows the soft X-ray count rates (0.4$-$2 keV) versus 
hardness ratio, where the hardness is defined  as 2$-$10 keV count rates 
divided by 0.4$-$2 keV count rates. This clearly suggests a spectral 
evolution such that the spectrum becomes softer when the source 
becomes brighter, but again the correlation is rather loose, 
especially when the source is in 
lower states of activity (obs $\#2 - 4$).  Also, it appears that  
the hardness diagram cannot be fit linearly against 0.4$-$2 keV 
source counts. This may suggest more than just one spectral 
component is responsible for the 3C~120 X-ray emission.

Similarly Figure \ref{fig:hd-pin} (a) 
compares the 2$-$10 keV XIS and 
12$-$40 keV  HXD/PIN count rates during the 3C~120 $\#1 - 4$
observations. Again, one can find a weak correlation between source 
variability in these two energy bands. In this particular case, 
however, the hardness seems to stay almost constant within statistical
errors as shown in Figure \ref{fig:hd-xis} (b). 
The apparent small changes in hardness ratio may suggest 
a relative lack of spectral evolution above 2 keV,  though hardness 
changes of $\sim$20 $\%$ level (as observed in Figure 
\ref{fig:hd-xis} (a)) are difficult to detect above 10 keV due 
to limited photon statistics. 

\subsection{Excess Variance}

For more detailed temporal studies, obs $\#1 - 4$ 
light curves are further divided into 6 energy bands 
(0.4$-$1 keV, 1$-2$ keV, 2$-$3 keV, 3$-$5
keV, 5$-$7 keV, 7$-$10 keV) for the XISs and 1 energy bands (12$-$40
keV) for the HXD/PIN, respectively. 
To estimate the amplitude of variability in a systematic way, 
``excess variance'' (e.g., Zhang et al.\ 2002 and reference therein) 
was calculated for light curves derived in different energy bands. 
Excess variance ($\sigma_{\rm rms}^2$), is a net variance, 
which is defined as the difference  between total variance 
($\sigma_{\rm tot}^2$) and noise variance ($\sigma_{\rm noise}^2$), 
that is caused by statistical errors;
\begin{equation}
\sigma_{\rm rms}^2 = \frac{1}{N \overline{x}^2}\sum_{i=1}^{N} [(x_i - \overline{x})^2 - \sigma_i^2] 
                   = \frac{1}{\overline{x}^2}[\sigma_{\rm tot}^2 - \sigma_{\rm noise}^2], 
\end{equation}
where $x_i$ is the $i$-th bin in the light curve and $\overline{x}$ is
the mean of $x_i$. 
The error on $\sigma_{\rm rms}^2$ is estimated by
$s_D$/($\overline{x}^2$$\sqrt{N}$), where $s_D$ is the variance of
quantity $(x_i - \overline{x})^2 - \sigma_i^2$ and given by
(Turner et al.\ 1999)
\begin{equation}
s_D^2 = 
\frac{1}{N - 1}\sum_{i=1}^{N} \{ [(x_i - \overline{x})^2 - 
 \sigma_i^2]  - \sigma_{\rm rms}^2 \overline{x}^2 \}^2.
\end{equation}
The fractional variability parameter F$_{\rm var}$ 
used in this paper is the square root of 
excess variance: F$_{\rm var} = (\sigma_{\rm rms}^2)^{1/2}$.

\begin{figure}
  \begin{center}
    \FigureFile(85mm,85mm){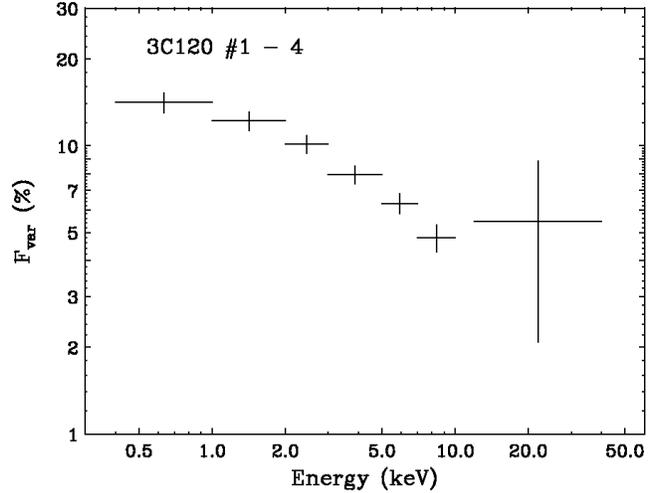}
  \end{center}
  \caption{Energy dependence of variability of 3C120 $\# 1 - 4$.
The variability parameter, excess variance, was calculated for the total
exposure in 7 (6 for 4 XISs + 1 for HXD/PIN) energy bands. The HXD/PIN data 
where the background is high (photon counts of $\ge$ 0.7 counts
 s$^{-1}$) are not used to minimize uncertainty due to background 
subtraction (see Appendix).
}\label{fig:exc1}
\end{figure}

Figure \ref{fig:exc1} shows  F$_{\rm var}$ measured in this way, 
using the overall light curves combined in obs $\#1 - 4$.  
Again, we carefully consider the effect of time dependent 
degradation of the XIS efficiency, but it only accounts  for $\sim$2 
$\%$ decrease of the XIS count rates even in 0.4$-$1.0 keV band. 
Clearly, the variability is larger in the lower energy bands; the largest 
variability was observed in the 0.4$-$ 1keV band with F$_{\rm var}$ = 
14.1$\pm$1.1 $\%$, and it gradually decreases with increasing energy, 
and reaches 4.8$\pm$0.5 $\%$ in 7$-$10 keV. The variability amplitude 
above 10 keV cannot be well constrained
(F$_{\rm var}$ = 5.5$\pm$3.4 $\%$), 
but is consistent with those of 7$-$10 keV variability observed with the XIS. 
Note this range of variability is consistent with that reported 
in Markowitz \& Edelson (2004), who find F$_{\rm var}$ = 6 $\%$ on 
6-day time scales and 8 $\%$ on 36-day time scale in 2$-$12 keV. 
We also calculated the excess variance of the 
light curves accumulated within each of the 40 ksec observations 
(obs $\#$1, 2, 3, 4, respectively).
The amplitude is smaller on this shorter time scale, 
with F$_{\rm var}$ $\le$ 3 $\%$ in all energy bands, and sometimes 
is consistent with no variability.
We therefore conclude that the variability time scale of 3C~120 is 
much larger than one day, typically $t_{\rm var}$ $\sim$ 10$^6$ (sec).
This relatively long timescale of variability is consistent with those 
found in literature (e.g., Marshall et al.\ 2003; 
Gliozzi et al.\ 2003; also see the Appendix for 
apparent short-term variations seen in the HXD/PIN light curve).

\section{Spectral Studies}
\subsection{Overview of the Broad-band Spectrum}

Before going into the detailed spectral study, we quickly overview  
the time-averaged spectrum of 3C~120 between 0.6 and 50 keV, from the whole 
Suzaku observation (obs $\#1 - 4$ combined). 
The XIS and HXD/PIN background subtracted spectra were fitted using 
XSPEC v11.3.2p, including data over the energy range 0.6$-$50 keV.  
The Galactic absorption toward
3C~120 is taken to be $N_{\rm H}$ =  1.23$\times$10$^{21}$ cm$^{-2}$ 
(Elvis, Wilkes \& Lockman 1989).  All errors are quoted at 68.3$\%$
(1$\sigma$) confidence for one interesting parameter unless otherwise 
stated. All the fits in this section are restricted in the energy range of 
0.6$-$12 keV (XIS 0,2.3: FI chips), 0.6$-$10 keV (XIS 1: BI chip), 
and  12$-$50 keV (HXD/PIN). 

\begin{table*}
  \caption{Results of spectral fits to the 3$-$50 keV XISs + PIN 
 time-averaged, co-added continuum spectrum of 3C~120.
 The 4 XISs data below 3keV (soft excess) and  between 5$-$7 keV 
(Fe-line complex) were discarded to avoid the complexity of the models.
}\label{tab:first}
  \begin{center}
    \begin{tabular}{lcccccccc}  
    \hline
    Model & $\Gamma$ & $E_{\rm fold}$ & $R$ & Abund & $i$ 
    & $F_{\rm 3-10keV}$$^b$ &  $F_{\rm 10-50keV}$$^c$ &  $\chi^2$/d.o.f\\
    & & [keV]  & [$\Omega$/2$\pi$]   & &  [deg] & & &  \\
    \hline\hline
    PRV$^a$   & 1.73$^{+0.03}_{-0.02}$ & 100$^f$ &
     0.47$^{+0.27}_{-0.12}$ & 1.0$^f$ & 18.2$^f$ & 
     30.0$\pm$0.1 & 62.1$\pm$0.6 & 1112/1060 \\  
    \hline
    \end{tabular}
  \end{center}
\small{$^a$ \textsc{pexrav} model in XSPEC 
(agdziarz \& Zdziarski 1995). We fixed the 
cutoff power-law energy $E_{\rm fold}$ = 100 keV and iron abundance
 $A_{\rm Fe}$ = 1.0.}\\
\small{$^b$ 3$-$10 keV flux in units of 10$^{-12}$ erg cm$^{-2}$ s$^{-1}$.}\\
\small{$^c$ 10$-$50 keV flux in units of 10$^{-12}$ erg cm$^{-2}$ s$^{-1}$.
       Constant fraction factor between HXD and XIS was set to be 
       $N_{\rm PIN}/N_{\rm XIS}$ = 1.114$^{+0.042}_{-0.079}$ 
       (see also Figure \ref{fig:cont}).}\\
\small{$^f$ Parameters fixed to these values.}\\
\end{table*}

\begin{figure}
  \begin{center}
    \FigureFile(85mm,85mm){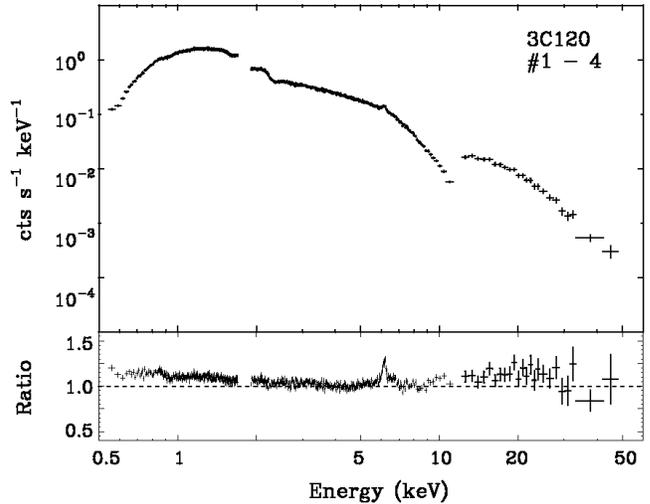}
  \end{center}
  \caption{The broad-band (0.6$-$50 keV; 3 FI-XISs + PIN) 
Suzaku spectrum of 3C~120.
The upper panel shows the data, plotted against an absorbed
 power-law model of photon index $\Gamma$ = 1.74 and column density 
1.23$\times$10$^{21}$ cm$^{-2}$, fitted over the 4$-$12 keV
 band. The lower panel shows the data/model ratio residuals to 
this power-law fit. Deviations due to (1) the iron  K-shell band, 
(2) the Compton reflection hump, and (3) excess soft emission are clearly 
seen.}\label{fig:multi-pw}
\end{figure}

As a first step, we fitted the 4 XIS spectra using a simple absorbed power-law 
model in the 3$-$10 keV band excluding 5$-$7 keV, in order for the 
continuum parameters not to be affected by possible iron line features, 
absorption/excess features in the lower energy band, as well as the  
reflection component expected to be present above 10 keV. 
Since we found that the spectral 
parameters of the 4 XIS spectra are all in good agreement within 
the error bars (at the few percent level), data from 3 FI-XISs
(XIS 0, 2, 3) were co-added to maximize the signal to noise. 
Figure \ref{fig:multi-pw} shows the 4 XISs + HXD/PIN 
spectra with residuals to this 
baseline model, where the XIS data around the Si K edge from 
1.7$-$1.9 keV were discarded in all 4 XIS chips, due to uncertain 
calibration at this energy at the time of this writing.
We also did not use the data below 0.6 keV, as uncertainties in 
C/O contamination are more significant (see $\S$2.1).  
\footnote{We note, an uncertainty in calibration affects 
only the ``spectral shape'' around the Si K edge. Similarly, the XIS 
spectra below 0.6 keV is still uncertain due to the C/O contamination, 
but no problems arises for the temporal studies as we have presented 
in $\S$3. } 

\begin{figure}
  \begin{center}
    \FigureFile(85mm,85mm){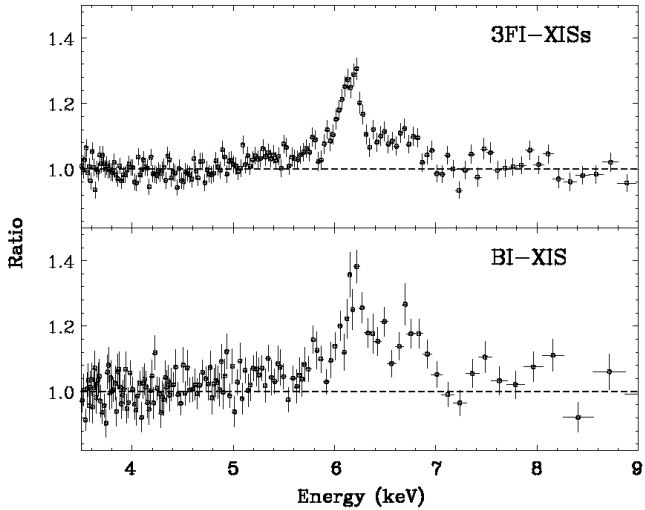}
  \end{center}
  \caption{A Zoom-in of the iron line profile of 3C~120, plotted as a
 ratio against a power-law of photon index $\Gamma$ = 1.74. ($upper$): XIS
 0,2,3 combined, ($lower$): XIS 1.}\label{fig:zoom-fe}
\end{figure}

The best fit spectral power-law index thus determined was 
$\Gamma$ = 1.74, but was statistically unacceptable if we 
extrapolate the model to lower and higher energies
 ($\chi^2$/d.o.f = 6593/614). The residuals of Figure \ref{fig:multi-pw}
 shows  that the spectrum exhibits features at 
different X-ray energies;  (1) iron line features around 6 keV,
(2) a hard X-ray bump above 10 keV, and 
(3) excess emission below 3 keV.  In particular, the structure 
around the iron K emission line is rather more complicated than 
that inferred from any previous X-ray satellite. A zoom-in of the 
``iron-line complex'' observed with Suzaku is presented 
respectively for the FI-XISs and BI-XIS in Figure \ref{fig:zoom-fe}.
Note the asymmetric line profile and the presence of a red-tail below 6 keV.

\subsection{The Baseline Continuum Emission}
 
To model the overall X-ray spectrum between 0.6 and 50 keV, we 
start from the shape of the baseline continuum emission. As for 
many radio-quiet AGNs, it has been suggested that the 
continuum of 3C~120 is composed of a direct (primary) power-law 
plus reflection by cold matter surrounding the nucleus 
(e.g., Wo\'{z}niak et al.\ 1998; Zdziarski \& Grandi 2001). 
We assumed a  primary component including an exponential 
cut-off at high energies of the form $\propto$ 
$E^{-\Gamma}$ $\times$ $e^{-E/E_{\rm fold}}$, where 
$E_{\rm fold}$ is the cut-off energy in keV, $\Gamma$ is 
the differential spectral photon index. 
Furthermore a reflection component produced by Compton 
down-scattering of X-rays off neutral material was included, using the 
\textsc{pexrav} model within XSPEC (Magdziarz \& Zdziarski 1995).
The inclination angle $i$ was fixed at 18.2 deg, which is the 
minimum allowed by the models, and also close to the limit 
derived from the superluminal motion of jet; see $\S$ 1. 
The solid angle $\Omega$ subtended by the Compton reflector was 
allowed to vary, and determined by the parameter $R$ = $\Omega$/2$\pi$.

In this modeling, we fixed the Fe abundance $A_{\rm Fe}$ 
at unity, following approaches previously made for this objects 
in literature (e.g., Ballantyne et al.\ 2004; Ogle et al.\ 2005). 
Similarly, we assumed a fixed value of $E_{\rm fold}$ = 100 keV because 
the HXD/PIN is not sensitive above 60 keV, and therefore
cannot determine the high energy end of direct power-law component 
(in this context, see Wo\'{z}niak et al.\ 1998 who confirmed a cutoff 
with $E_{\rm fold}$ $\sim$ 100 keV at a significance 
of 99.95$\%$ by combined ASCA and OSSE analysis. Ballantyne et al.\ 
(2004) assumed a fixed $E_{\rm fold}$ of 150 keV to fit a combined 
XMM-Newton and RXTE/HEXTE spectrum. Also Zdziarski \& Grandi (2000) found 
$E_{\rm fold}$ $\sim$ 120$-$150 keV using the Beppo-SAX data).  

As a first step, we fit the combined XIS + HXD/PIN data only above 3
keV, as the data below this energy is significantly contaminated by the 
soft excess emission as we have shown in Figure \ref{fig:multi-pw}. 
Similarly, data between 5 and 7 keV 
was ignored to avoid the region around the iron line complex. Even with this 
simplified model and constrained parameters, special care must be 
paid to the cross calibration of the XIS and the HXD/PIN, because 
it could easily affect the determination of the reflection parameter $R$. 
At the date of writing (2006 November), 
the relative normalization of the HXD/PIN 
was reported to be 1.13 times larger than the XIS, as determined 
by the XRT team using the spectrum of the Crab nebula 
(M.\ Ishida; JX-ISAS-Suzaku-memo 2006-40).

\begin{figure}
  \begin{center}
    \FigureFile(85mm,85mm){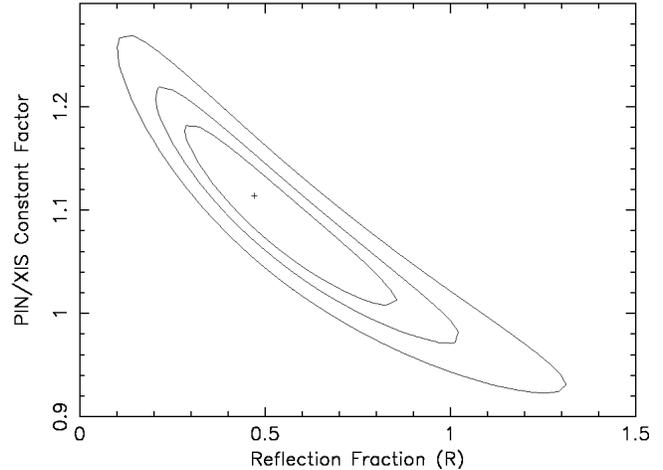}
  \end{center}
  \caption{Contour plot showing the reflection parameter $R$ versus the 
constant normalization factor between the HXD/PIN and XIS, which shows
 that the constant factor (defined here as $N_{\rm HXD/PIN}$/$N_{\rm XIS}$, 
where $N$ is the normalization) is close to 1.1, which is consistent
 with the cross calibration  between the XIS and the HXD 
using the Crab nebula. 
Confidence levels in the figure correspond to the 68$\%$,90$\%$ 
and  99$\%$ levels of significance, 
respectively.}\label{fig:cont}
\end{figure}

Figure \ref{fig:cont} shows the confidence contours of reflection parameter, 
$R$ = $\Omega$/2$\pi$, against the relative normalization of the
XISs/PIN. 
An apparent anti-correlation between the two parameters indicates $R$ is 
strongly affected by the normalization factor between the two instruments.
Nevertheless, the mean value of normalization factor 
(1.114$^{+0.042}_{-0.079}$) is perfectly consistent with that  
determined from the Crab nebula, and the reflection parameter is well 
constrained at $R$ = 0.47$^{+0.27}_{-0.12}$. The resultant parameters for the 
baseline continuum emission are summarized in 
Table 2. The photon index of the direct component is 
determined to be $\Gamma$ = 1.73$^{+0.03}_{-0.02}$. The 
observed flux  of the continuum emission (direct + reflection) is 
(30.0$\pm$0.1)$\times$10$^{-12}$ erg cm$^{-2}$ s$^{-1}$  over 3$-$10 keV, and 
(62.1$\pm$0.6)$\times$10$^{-12}$ erg cm$^{-2}$ s$^{-1}$  over 10$-$50 keV.

Finally, we comment on how the results are affected if we choose 
different values of $A_{\rm Fe}$ and $E_{\rm fold}$, which are
unfortunately  still uncertain even using the Suzaku data.
We are aware that fixing these parameters is an oversimplication
because both $A_{\rm Fe}$ and  $E_{\rm fold}$ could vary 
depending on the reflection parameter $R$.   
Using a fixed value of $E_{\rm fold}$ = 100 keV, both the 
reflection parameter and the power-law photon index are only slightly 
affected, where the best fit values provides 1.70 $\le$ $\Gamma$ $\le$ 
1.76 and 0.43 $\le$ $R$ $\le$ 0.53, respectively  for a range of 
0.5 $\le$ $A_{\rm Fe}$ $\le$ 2.0. If we set $E_{\rm fold}$ = 50 keV, 
reflection parameter becomes slightly large (0.77 $\le$ $R$ $\le$ 0.97 
for 0.5 $\le$ $A_{\rm Fe}$ $\le$ 2.0), but statistically not significantly 
different. The power-law photon index is within a range of 
1.70 $\le$ $\Gamma$ $\le$ 1.76.
Similarly, a different choice of $E_{\rm fold}$ = 150 keV provides  
0.37 $\le$ $R$ $\le$ 0.45 for 0.5 $\le$ $A_{\rm Fe}$ $\le$ 2.0, with 
1.72 $\le$ $\Gamma$ $\le$ 1.76. Again these are relatively small 
effects and thus we assume $A_{\rm Fe}$ = 1.0  and $E_{\rm fold}$ = 100 keV 
in the following analysis.

\subsection{Iron Line Complex}
\subsubsection{(1) Line Profile}

\begin{figure}
  \begin{center}
    \FigureFile(85mm,85mm){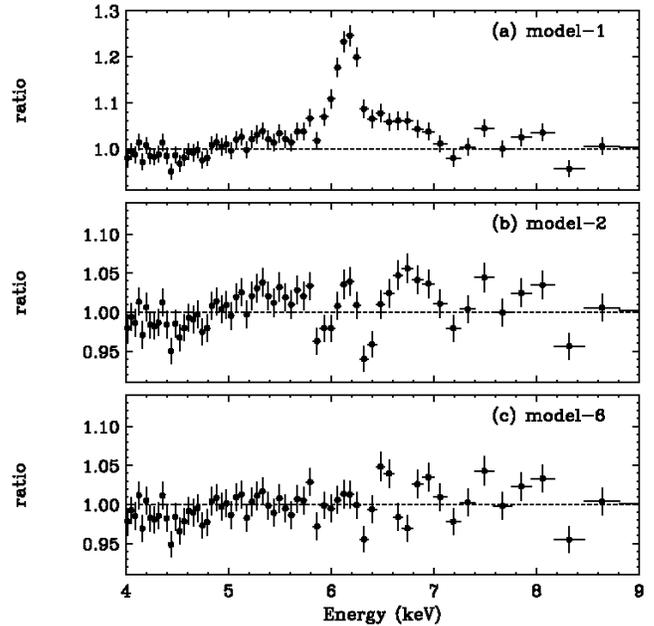}
  \end{center}
  \caption{Data/model ratio residuals of the 3 FI-XISs spectrum of 
3C~120 between 4$-$9 keV. $upper$ (a): residual after fitting a power-law
 ($\Gamma$=1.73) with a reflection component ($R$ = 0.47: model 1 in 
Table 3). 
$middle$ (b): residuals after inclusion of the narrow K$_{\alpha}$ core 
(model 2 in Table 3), $bottom$ (c): residuals after the addition of a 
broad diskline emission centered at 6.4 keV, as well as a 6.9 keV line 
(model 6 in Table 3).}\label{fig:Fe-res}
\end{figure}

We next consider the X-ray spectrum of 3C~120 between 3 and 10 
keV, with the inclusion of the 5$-$7 keV data. 
The iron K line profile was then fitted in several steps. 
Firstly we fit the joint FI-XISs (XIS 0,2,3) 
and BI-XIS (XIS 1) hard X-ray spectra with the best-fit power-law plus  
reflection model (\textsc{pexrav}: PRV) 
as described in the previous section. The fit is 
very poor as shown in Figure \ref{fig:Fe-res} 
(a) and Table.3 ($\chi^2_\nu$ =
2500/1606 = 1.56; model-1 in Table 3), mostly due to a prominent 
line profile near 6.2 keV in the observed frame 
($\sim$6.4 keV in the rest frame). Then, adding a single 
Gaussian Fe K$_\alpha$ line to the model (PRV+G) gives a greatly
improved fit statistic ($\chi^2_\nu$ = 1807/1603 = 1.13; model-2). 
The line energy, converted into the rest frame, is
6.378$^{+0.013}_{-0.009}$ keV, indicating neutral or low-ionization Fe. 
The measured width of the line is $\sigma_{\rm obs}$ = 193$^{+32}_{-31}$
eV, and the equivalent width is EW = 101$^{+17}_{-16}$ eV. For completeness, 
we also added a small Compton shoulder to the iron K$_{\alpha}$ line, 
represented by a narrow Gaussian centered at 6.24 keV, with normalization
fixed to 20$\%$ of the K$_{\alpha}$ flux (Matt 2000), but 
there was no improvement to the fit at 90$\%$ confidence (no significant improvement
in $\chi^2$). 

\begin{table*}
  \caption{Results of the spectral fits to the 3$-$10 keV XIS 0$-$3
 time-averaged, co-added spectrum with the different models used to
 describe the Fe-line complex of 3C~120.}\label{tab:first}
  \begin{center}
    \begin{tabular}{lccccccccc}  
    \hline
    Model & E$_1$  & $\sigma_{\rm 1}^d$ 
          & E$_2$  & $\sigma_{\rm 2}^d$
          & E$_{\rm DL}$  & $r_{\rm in}$ & $i$ & $\chi^2$\\
          & [keV]  & (EW) [eV] & [keV] & (EW) [eV] & [keV] 
          & [$r_g$] & [deg]  & (d.o.f)  \\
    \hline\hline
    1. PRV$^a$    & ... & ... & ... & ... & ... & ... & ...  & 2500 \\  
          & & & & & & & & (1606)\\ 
    2. PRV$^a$+G$^b$ & 6.378$^{+0.013}_{-0.009}$ & 193$^{+32}_{-31}$ & ...
     & ... & ... & ...  & ... & 1807 \\  
          & & (101$^{+17}_{-16}$) &  & & & & & (1603)\\
    3. PRV+DL$^c$ & ... & ...  & .. &  ...
     & 6.390$^{+0.007}_{-0.009}$  & 40.7$^{+6.9}_{-5.0}$ & 18.2$^f$ & 1853\\
          &  &  &  &  &  & &  & (1603)\\
    4. PRV+2G & 6.373$\pm$0.009 & 158$^{+15}_{-17}$ &
     6.93$\pm$0.02 & 10$^f$ & ... & ... & ... & 1757 \\  
          & & (90$^{+9}_{-10}$) &  & (18) & & & & (1601)\\
    5. PRV+G+DL & 6.368$^{+0.008}_{-0.006}$ & 69$^{+6}_{-8}$ & ... & ...
     & 6.41$^{+0.10}_{-0.06}$ &  19.4$^{+5.1}_{-1.8}$ & 49.7$^{+1.9}_{-7.1}$  & 1691 \\  
          & & (49$^{+5}_{-6}$) & & & & & & (1599)\\ 
    6. PRV+2G+DL & 6.398$\pm$0.01 & 111$^{+11}_{-10}$ & 6.93$\pm$0.02 &
     10$^f$ & 6.40$^f$ &  8.6$^{+1.0}_{-0.6}$ & 6.5$^{+3.8}_{-3.2}$  & 1681 \\  
          & & (60$\pm$6) & & (17) & &   & & (1598)\\ 
    \hline
    \end{tabular}
  \end{center}
\small{$^a$ \textsc{pexrav} model in XSPEC (Magdziarz $\&$ Zdziarski 1995).
We assumed the direct power-law of $\Gamma$ = 1.73  with 3$-$10 keV flux 
$F_{\rm 3-10 keV}$ = 3.0$\times$10$^{-11}$ erg cm$^{-2}$ sec$^{-1}$ (see
 Table 2). We fixed the reflection parameter $R$ = 0.47 and Fe abundance 
$A_{\rm Fe}$ = 1.0.}\\
\small{$^b$ A simple \textsc{gauss} function in XSPEC modified by redshift 
of 3C~120.}\\
\small{$^c$ \textsc{diskline} model assuming a power-law dependence of emissivity, $\beta$ = $-$3.0.}\\
\small{$^d$ An intrinsic width of iron line (excluding residual 
width in calibration $\sigma_{\rm cal}$ = 45$\pm$1 eV; see $\S$ 2.1).}\\
\small{$^e$ 3$-$10 keV flux in units of 10$^{-12}$ erg cm$^{-2}$
 s$^{-1}$.}\\
\small{$^f$ parameters fixed to these values.}\\
\end{table*}

After the inclusion of the iron K line, there still remains 
a clear excess of photon counts in the Suzaku data between 
5 and 7 keV (Figure \ref{fig:Fe-res} (b)).  We further tried to  
improve the fit by assuming a redshifted diskline rather 
than a simple Gaussian  function (PRV +DL: model-3), or by adding another 
Gaussian to the model (PRV + 2G; model-4).  
Indeed the latter model (model-4) is suggested to  fit the XMM-Newton data
pretty well (Ogle et al.\ 2005),  and also improves the $\chi^2$ 
statistic of the Suzaku data significantly ($\chi^2_\nu$ = 1757/1601 =
1.10). However the inclusion of an additional Gaussian still 
does not represent 
very well the ``red tail'' below 6 keV observed in the Suzaku data. 
Instead, inclusion 
of a broad diskline from an accretion disk (Fabian et al.\ 1989; 
see also Reeves et al.\ 2006 to fit the similar iron line 
profile of MCG--5-23-16) provides a better representation of the data. 
As a result, the fit statistic of our ``best-fit'' model (PRV+G+DL) was 
improved to $\chi^2_\nu$ = 1691/1599 = 1.06 (model-5). 
This ``PRV+G+DL'' model (model-5), however, predicts a large inclination angle
of $i$ $\simeq$ 50 deg. This seems problematic when taking into account the  
tight constraints on the jet inclination angle of $i$ $\le$ 14 deg from 
the superluminal motion, unless the jet is significantly
``warped'' between the disk and the VLBI (i.e., pc scale) region. 
Also note that the centroid energy of the Fe line core is 
6.368$^{+0.008}_{-0.006}$ keV, which is shifted redwards by $\sim30$\,eV 
compared to what is expected from a neutral iron K$_{\alpha}$ line 
(6.400 keV).  
This shift is effectively larger than the uncertainties in the energy scale 
determined by the calibration sources (see $\S$ 2.1).

We therefore considered whether we could fit the same line profile 
differently, to achieve a more face-on orientation. Basically,
the high inclination angle is driven from fitting the blue-wing 
of the line above 6.4 keV. If we assumed there is ionized emission 
from either He-like (6.7 keV) or H-like (6.97 keV) Fe K$_{\alpha}$
as was suggested in the Chandra HETGS and the XMM-Newton observations, 
that would reduce the inclination, 
as the blue-wing of the diskline would not need 
to extend much beyond 6.4 keV. While the fit with this ``PRV+2G+DL''
model (model-6 in Table 3) is only marginally better than before, the 
diskline now has a reasonable inclination of $i$ = 6.5$^{+3.8}_{-3.2}$ 
deg with an EW of 32$\pm$5 eV, and fits the red-wing of the line profile well 
(while the ionized Fe emission models the blue-wing: 
see Figure \ref{fig:Fe-res}(c)). 
It is also interesting to note that the centroid energy of 
the narrow Fe line core increases 
from 6.368 keV to 6.398 keV (perfectly consistent with 6.400 keV 
within a 1$\sigma$ error) in this face-on model ($i$ $\le$ 10 deg). 

The measured width of narrow Fe line core ($\sigma_{\rm obs}$ $\sim$120
eV) is much broader than the residual width ($\sigma_{\rm cal}$) of a 
calibration line.  Assuming $\sigma_{\rm cal}$ = 45 eV, the intrinsic 
width of Fe K$_{\alpha}$ line is $\sigma_{\rm int}$ = 111$^{+11}_{-10}$ eV 
(or EW of 60$\pm$6 eV). This line width is consistent with what 
was measured with XMM-Newton very recently by Ogle et al.\ (2005) and 
Ballantyne et al.\ (2004). Meanwhile, the additional broad disk line 
emission provides a better representation of the data if we fixed the center 
energy at 6.4 keV. Here we have assumed the outer radius 
of the diskline is $r_{\rm out}$ = 1000 $r_g$, and a typical steep 
emissivity of $r^{-3}$. The inner radius of the disk is constrained to be  
$r_{\rm in}$ = 8.6$^{+1.0}_{-0.6}$ $r_g$. A summary of the line
parameters is given in Table 3. 

\subsubsection{(2) Line Variability}

We next consider the iron line variability and its relation to the variation 
of the baseline continuum component. Due to limited photon statistics, 
we traced the variability of the Fe K$_{\alpha}$ line core only, 
using a fixed line width and centroid energy.  For this, a simple  power-law 
plus Gaussian model is adequate to represent the Suzaku data in the 3$-$10 
keV band. Thus in our approximation, the only parameter which represents  
line variability is its normalization (i.e., integrated photon counts in 
iron line core) measured in different observation epochs. We are 
aware this is an oversimplified assumption given the complicated line 
profile and that variations with time are even possible for the 
reflection component. Nevertheless, we believe this is the simplest 
way of searching for line variability on short timescales. 
The XIS data were divided into 4 equal time intervals 
(typically $\sim$10 ksec) per each $\#1 - 4$ observations. Even 
assuming this simple PL + G model, all the datasets provide acceptable fits 
in the statistical sense of $P(\chi^2)$ $\ge$ 10$\%$. 

\begin{figure}
  \begin{center}
    \FigureFile(85mm,85mm){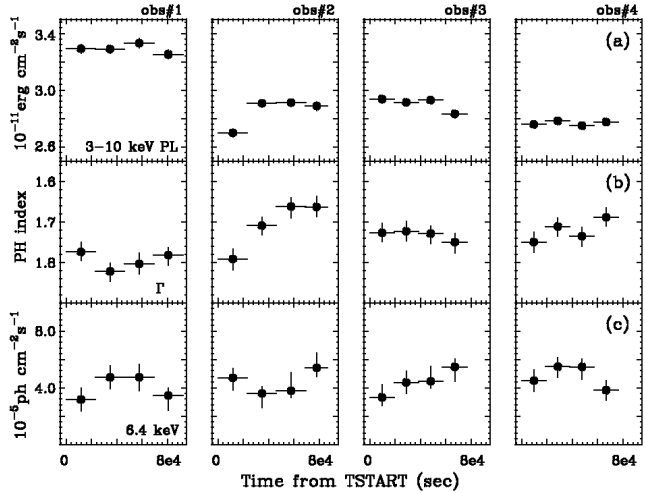}
  \end{center}
  \caption{Spectral variability of baseline power-law component
 and iron K$_{\alpha}$ core emission. 3$-$10 keV flux ($upper$), photon
 index ($middle$), and line flux ($bottom$) are separately shown. 4 XISs 
 data are used for the fitting.
}\label{fig:line-var1}
\end{figure}

\begin{figure}
  \begin{center}
    \FigureFile(85mm,85mm){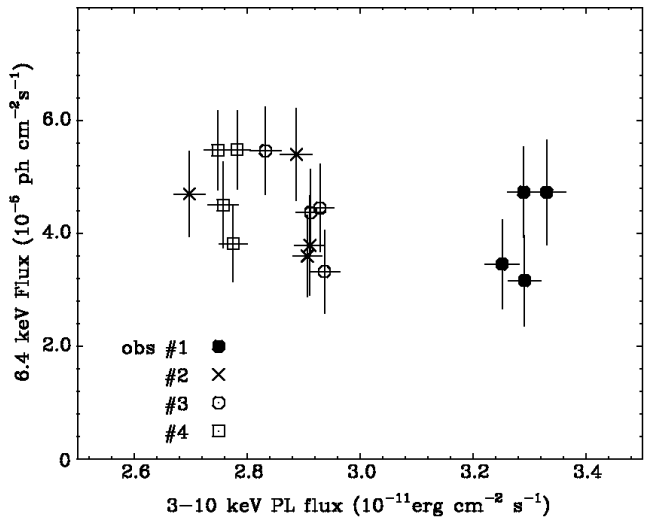}
  \end{center}
  \caption{Comparison of the iron K$_{\alpha}$ core flux versus baseline
 continuum component. Variability is not clear for the iron line flux.
}\label{fig:line-var2}
\end{figure}

Figure \ref{fig:line-var1} shows the spectral variability of the 
baseline power-law component
and the iron K$_{\alpha}$ core emission. Temporal evolution of 
3$-$10 keV continuum flux, photon index, and Fe line core flux are 
plotted in the upper, middle, 
and bottom panels, respectively. Although  variability at some level 
may be present in the light curve of Fe line core flux,  a constant fit 
provides a good explanation of the data, where $\chi^2$/d.o.f = 17.3/15 
(P($\chi^2$) = 30$\%$) and an average iron line photon flux of  $F_{\alpha}$ = 
(4.41$\pm$0.19)$\times$10$^{-5}$ ph cm$^{-2}$ s$^{-1}$.  
A correlation between the underlying power-law continuum 
and Fe line fluxes is uncertain, but seems to be absent as shown 
in Figure \ref{fig:line-var2}. These results, however, do not 
necessarily reject any variations of the iron line profile itself, 
if the intensity has not changed significantly during the observation.  
In fact, the iron line profile observed with Suzaku seems to have 
a somewhat different shape than that observed with the XMM-Newton in 2003.
It seems that only the Suzaku data requires a strong red-tail, but 
instead the 6.9 keV emission seems to be a little more ``spiky'' in 
the XMM-Newton data (but we need more photon statistics to discriminate this 
further). 
A further detailed analysis of the line/absorption features is now on-going
and will be presented in a forthcoming paper, especially by 
direct comparison of the XMM-Newton and the Suzaku spectra 
(Iwasawa et al.\ in prep). 

\subsection{Excess Soft X-ray Emission}

Finally we try to model the excess soft X-ray emission below 3 keV, 
as presented in Figure \ref{fig:multi-pw}. We found both a 
steep power-law ($\Gamma$ = 2.68$\pm$0.04) or thin thermal 
bremss emission ($kT$ = 1.06$\pm$0.04 keV) can equally well fit the data 
with $\chi^2$/dof $\simeq$ 1.1. Also the inclusion of an emission line 
feature near 0.9 keV improves the fit slightly, though it might an 
artifact of modeling the XIS data around Ne I K edge (see Ogle 
et al.\ 2005 for XMM-Newton RGS soft X-ray spectrum). 
Interestingly the excess emission below 3 keV seems a bit  
$smaller$ in this Suzaku 
spectrum than that reported in the XMM-Newton observation 
(Figure 2 of Ballantyne et al.\ 2005), for which the source was in a 
brighter state than any of the Suzaku observations (obs $\#1 - 4$).
Such excess emission was $not$ necessary in the ASCA analysis 
(Grandi et al.\ 1997), but was found 
in both the ROSAT (Grandi et al.\ 1997) and BeppoSAX data 
(Zdziarski \& Grandi 2001), as well as in the XMM-Newton data.
 
This kind of soft excess emission has been observed in 
other BLRGs (Wo\'{z}niak et al.\ 1998) and Sy-1 galaxies, and its 
nature is currently under debate.  For 3C~120, it was first 
discovered in the Einstein SSS observations (Petre et al. 1984).
Grandi et al.\ (1997) fit the 0.2$-$2.4 keV ROSAT data and obtained 
the best fit results with a steep power-law emission of 
$\Gamma$ = 2.5$-$3.3, whereas Zdziarski \& Grandi favor thin 
thermal emission of $kT$ $\sim$ 1 keV. 
Very recently, Ogle et al (2005) suggested a steep power-law with $\Gamma$ = 
2.7$\pm$0.1, which breaks at 0.6 keV to represent the XMM-Newton data below 1
keV. In contrast Ballantyne et al.\ (2004) assumed a 
thermal bremsstrahlung component in order to reproduce the same X-ray data. Therefore 
the situation is not still conclusive even after the XMM-Newton 
and the Suzaku observations. But if the soft excess is due to
collisionally ionized plasma in an extended halo surrounding the 
nucleus as suggested in Zdziarski \& Grandi 2001, this should vary very little 
(but see the discussion in Ballantyne et al.\ 2004).  

In fact, in some Seyfert galaxies, the soft excess emission is 
consistent with being constant and is likely to be a combination of 
scattered power-law emission and photoionized gas associated 
with the AGN Narrow Line Region etc (e.g., Vaughan \& Fabian (2004); 
Reeves et al.\ (2006)).
Statistically we cannot rule out there being some 
thermal/bremss emission, but we will discuss below the 
power-law continuum emission which makes more sense 
in the context of the two components required to 
explain the spectral variability (see $\S$ 4.5). Moreover, with the observed 
variability being more strongly concentrated at lower energies, a more natural 
interpretation appears to be a steep power-law component.  

\subsection{Multiband Spectral Variability}

\begin{figure}
  \begin{center}
    \FigureFile(85mm,85mm){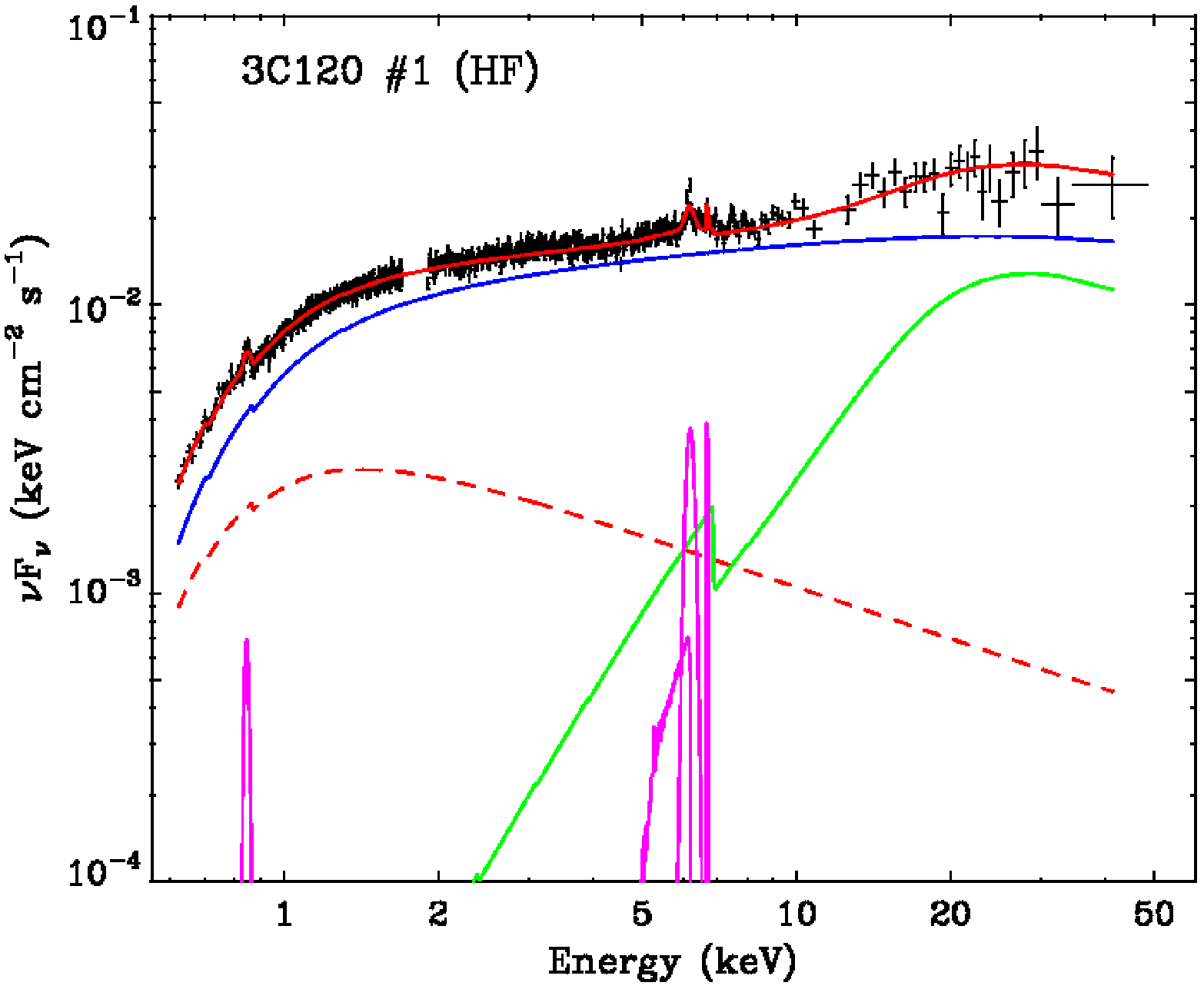}
    \FigureFile(85mm,85mm){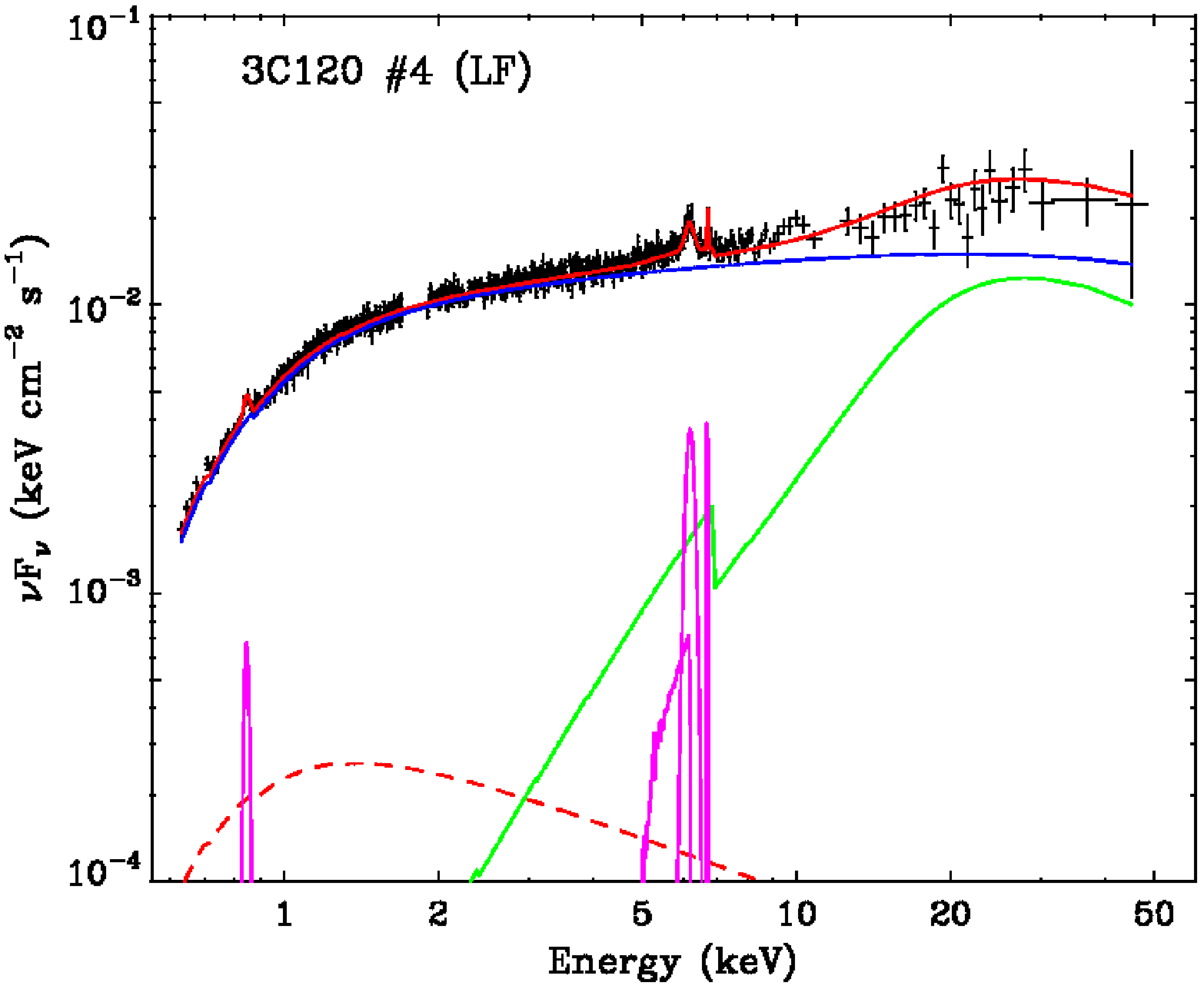}
  \end{center}
  \caption{Broad-band (0.6$-$50 keV: 3FI-XISs + PIN) 
spectra of 3C~120 observed with 
Suzaku in high ($upper$: obs $\#$1) and low ($lower$: obs $\#$4) states, 
with unfolded best-fit model components. 
$red$: total, $blue$: direct power-law component,  $green$: reflection, 
$purple$: iron lines, $red$ $dashes$; variable power-law component.  
The best fit parameters are summarized in Table 4.}
\label{fig:mulbest}
\end{figure}

\begin{figure}
  \begin{center}
    \FigureFile(85mm,85mm){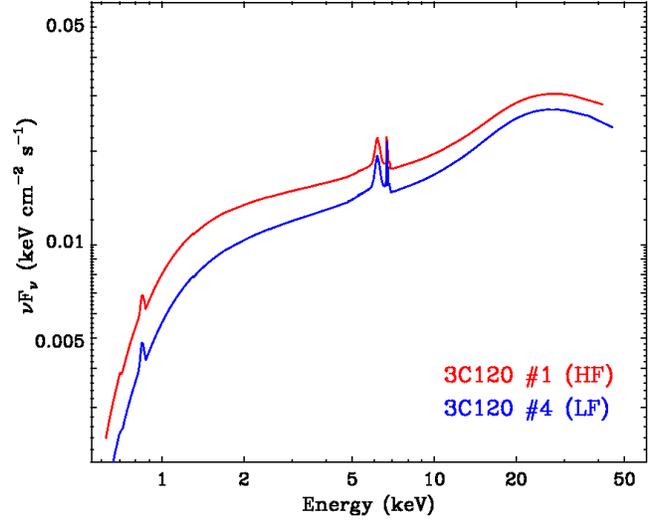}
  \end{center}
  \caption{Comparison of the best-fit broad-band (0.6$-$50 keV) model 
of 3C~120 observed with Suzaku in high flux (HF; $red$: obs $\#$1) and 
low flux 
 (LF; $blue$: obs $\#$4) states with the best-fit model functions. 
The best fit parameters for HF/LF states are summarized in Table 4.}\label{fig:mulcomp}
\end{figure}

Figure \ref{fig:mulbest} shows the unfolded multiband spectra
of 3C~120 obtained with Suzaku in both high (obs $\#$1) and low (obs $\#$4)
states. The best-fit model described in the previous sections 
is taken into account as various emission components in 
these figures; namely
a direct power-law, Compton reflection, a narrow Fe line core, a 6.9 keV
line,  a broad disk line, and a steep power-law emission to represent
the soft excess emission below 3 keV. We first assumed the values 
determined in the previous sections as initial input parameters, 
and then re-fit the data again to find a $\chi^2$ minimum under the 
constraint of fixed iron line parameters. This is because if we limit 
the analysis $only$ above or below 3 keV as we have done in $\S$4, 
this avoids having to include a scattered continuum component which 
could affect the fit. Therefore we thawed all the parameters 
except for iron line parameters, to find a ``new'' $\chi^2$ minimum.
The results of the fits are summarized in Table 4  
for various cases; (1) time-averaged spectra (obs $\#1 - 4$ summed), 
(2) high flux state (HF; obs $\#$1), and (3) low flux state (LF; obs $\#$4), 
respectively. Direct comparison of the multiband spectra between HF and 
LF states (Figure \ref{fig:mulcomp})  clearly indicates energy dependence 
of spectral evolution in different states of source activity.

\begin{table*}
  \caption{Results of spectral fits to the 0.6$-$50 keV XIS+PIN 
 time-averaged, co-added spectrum of 3C~120. 
}\label{tab:first}
  \begin{center}
    \begin{tabular}{llccc}  
    \hline
    Component & Parameter & Average & High (obs $\#1$) & Low (obs $\#$4)\\
    \hline
   wabs  & $N_{\rm H}$$^a$  & 1.23$^{f}$ & $-$$^d$ & $-$
     \\
    \hline  
    PRV  & $\Gamma$ & 1.74$\pm$0.02 & 1.76$\pm$0.03 & 1.78$\pm$0.03 \\  
         & $E_{\rm fold}$ [keV] & 100$^f$ & $-$ & $-$  \\
         & $R$ & 0.65$\pm$0.12  & 0.79$\pm$0.22 & 0.89$\pm$0.20 \\
         & $i$ [deg] & 18.2$^f$  & $-$ & $-$ \\    
         & $F_{\rm 0.6-3keV}$$^b$ & 26.2$\pm$0.5 & 28.6$\pm$1.5  & 26.5$\pm$1.0\\
         & $F_{\rm 3-10keV}$$^b$ & 28.7$\pm$0.2 & 30.5$\pm$0.5  & 27.6$\pm$0.3\\
         & $F_{\rm 10-50keV}$$^b$ & 63.3$\pm$0.8 & 68.4$\pm$1.5 & 61.8$\pm$1.3\\
    \hline
    Gauss1 & $E$ [keV] & 6.40$^f$ & $-$  & $-$  \\  
         & $\sigma$ [eV] & 111$^f$ & $-$   & $-$ \\ 
         & $N_{\rm line}$$^c$ & 3.14$^f$ & $-$ & $-$ \\ 
    \hline 
    Gauss2 & $E$ [keV] & 6.93$^f$ & $-$  & $-$  \\  
         & $\sigma$ [eV] & 10$^f$ & $-$   & $-$ \\ 
         & $N_{\rm line}$$^c$ & 0.69$^f$ & $-$ & $-$ \\ 
    \hline 
    Diskline  & $E$ [keV] & 6.40$^f$ & $-$  & $-$  \\  
         & $R_{\rm in}$ [r$_g$] & 8.6$^f$ & $-$  & $-$  \\    
         & $i$ [deg] & 6.5$^f$  & $-$  & $-$  \\  
         & $N_{\rm line}$$^c$ & 1.57$^f$ & $-$  & $-$  \\   
    \hline
    PL  & $\Gamma$ & 2.66$\pm$0.04 &  2.60$\pm$0.08 & 2.65$^f$   \\
        & $F_{\rm 0.6-3keV}$$^b$ & 4.8$\pm$0.6  & 9.3$\pm$1.8 & 1.1$^{+1.3}_{-1.1}$ \\
        & $F_{\rm 3-10keV}$$^b$ & 1.4$\pm$0.2  & 2.9$\pm$0.6 & 0.3$^{+0.4}_{-0.3}$ \\
    \hline
    Gauss3 & $E$ [keV] & 0.87$^f$ & $-$  & $-$  \\    
         & $\sigma$ [eV] & 10$^f$ & $-$  & $-$  \\   
         & $N_{\rm line}$$^c$ & 5.71$^f$ & $-$  & $-$  \\   
    \hline 
       $\chi^2$/d.o.f &  & 2978/2768 &  3018/2884  &  2853/2729   \\  
    \hline
    \end{tabular}
  \end{center}
\small{$^a$ Galactic absorption column density in units of 10$^{21}$ cm$^{-2}.$}\\
\small{$^b$ Flux in units of 10$^{-12}$ erg cm$^{-2}$ s$^{-1}$.}\\
\small{$^c$ Line normalization in units of 10$^{-5}$ photons cm$^{-2}$
 s$^{-1}$.}\\
\small{$^d$ Assumed to be same as the left, i.e., those assumed 
to fit an``average'' spectrum.}\\
\small{$^f$ Parameters fixed to these values.}\\
\end{table*}

Thus it is interesting to consider what spectral component is 
primarily responsible for the spectral variability in 3C~120. 
As can be seen in Table 4 and  Figure \ref{fig:mulbest}, 
the underlying continuum emission 
(i.e., the sum of the  direct power-law 
plus reflection) varies only a little.  
In fact, the spectral shapes, as measured by a photon index of $\Gamma$
$\simeq$ 1.75 in LF and HF states, are consistent within the statistical
error. The flux actually changes between LF and HF states, but only at the 10$\%$ level between 0.6$-$50 keV. In contrast, a factor of $\ge$5 variability has 
been observed for the steep power-law component where the 0.6$-$3 keV flux 
changes significantly from (9.3$\pm$1.8)$\times$$10^{-12}$ erg 
cm$^{-2}$ s$^{-1}$ to (1.1$^{+1.3}_{-1.1}$)$\times$10$^{-12}$ erg 
cm$^{-2}$ s$^{-1}$. Actually, this steep power-law component is 
$not$ visible when the source is in a low flux state, and only appears 
when the source is in a brighter state. This is consistent with the fact 
that the XMM-Newton spectrum shows a slightly larger amount of soft excess emission 
when the source was in a relatively bright flux state compared to 
any of the Suzaku observations described in this paper.

\begin{table*}
  \caption{Results of spectral fits to the 0.3$-$50 keV $difference$ 
spectrum between obs $\#$1 and obs $\#$4.
}\label{tab:first}
  \begin{center}
    \begin{tabular}{lcccccccc}  
    \hline
    Model$^a$ & $kT$ [keV] & $F_{\rm brm}^b$  & $\Gamma_1^c$ & $F_{\rm PL1}^b$ & $\Gamma_2^c$  & $F_{\rm PL2}^b$ & $\chi^2$/d.o.f\\
    \hline\hline
    wabs + Bremss  & 2.53$\pm$0.07 & 13.0$\pm$0.5  & ... & ...  & ... & ... & 453/192 \\  
    wabs + PL & ... & ... & 2.22$\pm$0.02 & 14.8$\pm$0.2 & ...  & ... & 198/192 \\  
    wabs + 2PL & ...& ... & 2.65$^f$ & 9.4$\pm$0.3  & 1.75$^f$ & 6.3$\pm$0.3  &  198/192 \\  
    \hline
    \end{tabular}
  \end{center}
\small{$^a$ A thin-thermal or power-law function modified by Galactic
 absorption of $N_H$ = 1.23$\times$10$^{23}$ cm$^2$.}\\
\small{$^b$ 0.6$-$10 keV flux in units of 10$^{-12}$ erg cm$^{-2}$ s$^{-1}$}\\
\small{$^c$ Differential spectral photon index.}\\
\small{$^f$ Parameters fixed to these values.}\\
\end{table*}
  
\begin{figure}
  \begin{center}
    \FigureFile(85mm,85mm){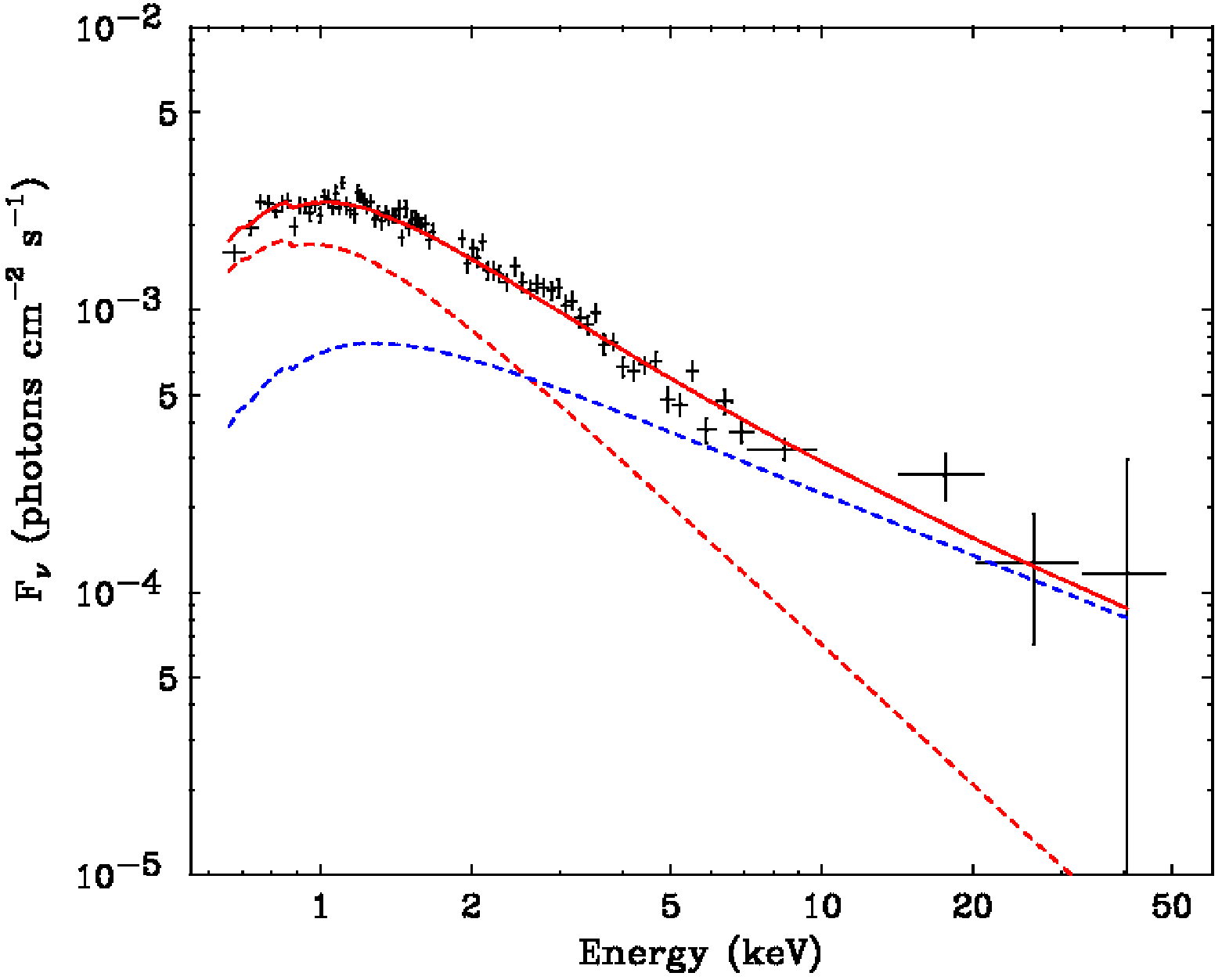}
  \end{center}
  \caption{The difference spectrum of 3C~120 between high (obs $\#$1) 
and low (obs $\#$4) flux  states in the Suzaku observation (3 FI-XISs +
 PIN). Although 
the difference spectrum is well represented by an absorbed power-law 
of $\Gamma$ = 2.22$\pm$0.02 in the XIS energy band, some residuals 
remains above 10 keV.  The overall spectrum (including HXD/PIN) 
is better represented by a double power-law function, where the 
low energy emission (below 3 keV)  is dominated by a steep power-law of 
$\Gamma$ = 2.65 ($red$ $dashes$) and the high energy emission (above 10 keV) 
is modified by a flat power-law of $\Gamma$ = 1.75 ($blue$ $dashes$). }\label{fig:diff}
\end{figure}

To further examine the origin of the spectral variability, 
the difference spectrum of HF minus LF spectra was extracted, using 
both the XIS and HXD/PIN data. The difference spectrum is plotted in 
Figure \ref{fig:diff}. This technique clearly  shows the 
variable component of the emission from 3C~120 modified by absorption, 
with the constant components in the spectrum being subtracted. 
The resulting difference spectrum was fitted very well with a 
simple power-law of photon index $\Gamma$ = 2.22$\pm$0.02 and 
fixed $N_{\rm H}$ = 1.23$\times$10$^{21}$ cm$^{-2}$, with the fit 
statistic of $\chi^2$/dof = 198/192. Note that the inclusion of a thin 
thermal emission component provides a rather poor fit as shown in Table 5.

 Although this spectral shape is a bit flatter than a steep power-law 
describing the soft excess emission ($\Gamma$ $\simeq$ 2.7; 
see PL of Table 4), only small variations in the underlying 
continuum emission (PRV: $\Gamma$ $\simeq$ 1.7) can easily 
account for this apparent discrepancy.  
In fact, the same difference spectrum can be equally well fitted 
with double power-law functions, where the low energy emission 
is dominated by a steep power-law ($\Gamma_1$ = 2.65, which mimics the soft
``variable'' power-law) and the high energy emission 
is represented by a flat power-law ($\Gamma_2$ = 1.75, which mimics the hard 
``direct'' power-law component). These model components are presented 
in Figure \ref{fig:diff} as red and blue dashes, respectively. 

We also note that a normalization of the HXD/PIN difference 
spectrum seems a bit larger than an extrapolation of the XIS 
spectrum above 10 keV. This may imply both intrinsic and reprocessed 
component change together in 3C~120, though it was not the case for 
MCG-6-30-15, where the reflected component is constant and 
only the intrinsic power-law varies (Minuitti et al.\ 2006: see
also $\S$5.3).
But we suspect a part of the reason could be instrumental: 
As we see in $\S$2 and in the Appendix, current background 
models of the HXD/PIN would produce $\sim$5$\%$ systematics in the 
background subtraction, which corresponds to $\sim$20$\%$ 
uncertainties in the 3C~120 flux above 10 keV. This could be more 
enhanced in the differnce spectrum as shown  in Figure \ref{fig:diff}. 
We will discuss more about this in a forthcoming paper using update
background models of the HXD/PIN.
 
\section{Discussion}
\subsection{Suzaku View of 3C~120}

In previous sections, we presented temporal and spectral analysis of 
a Suzaku observation of 3C~120 in February and March 2006. The great advantage 
of Suzaku is that both the XIS and HXD/PIN have excellent sensitivity, 
so that we can resolve the spectral evolution of this source 
over two decades in frequency even within a short (40 ksec) exposure. 
During the Suzaku observations, 3C~120 was in a relatively low flux state of 
$F_{\rm 2-10 keV}$ = 3.9$\times$10$^{-11}$ erg cm$^{-2}$ s$^{-1}$ on average, 
though the flux decreased gradually from 
4.4$\times$10$^{-11}$ erg cm$^{-2}$ s$^{-1}$ to 
3.6$\times$10$^{-11}$ erg cm$^{-2}$ s$^{-1}$ between obs $\#$1 and obs $\#$4. 
Historically, the maximum flux observed by Suzaku 
(that obtained in obs $\#1$) is almost consistent with what was observed 
by ASCA in 1994,  by Beppo-SAX in 1997 and by XMM-Newton in 2003 
($F_{\rm 2-10 keV}$ $\sim$ 4.6$\times$10$^{-11}$ erg cm$^{-2}$ s$^{-1}$; 
Reynolds 1997; Zdziarski \& Grandi 2001; Ballantyne et al.\ 2004) 
but 40 $\%$ lower than that reported by RXTE in 1998 
(Eracleous et al.\ 2000). This range of variability is natural 
considering the long term variability of this source over 
year-long timescales in X-rays (Marscher et al.\ 2002; Markowitz \&
Edelson 2004).

Using the high sensitivity, broad-band instruments onboard Suzaku, 
we have confirmed various important characteristics of 3C~120 ; 
(1) the presence of a relatively narrow ($\sigma$ = 111$^{+11}_{-10}$
eV or EW of 60$\pm$6 eV) Fe line core centered at 6.4 keV 
(see Ballantyne et al.\ 2004; 
Ogle et al.\ 2005 for XMM-Newton analysis). In addition, Suzaku has 
unambiguously confirmed the broad iron line component which may be possibly 
related with an extremely broad line feature observed with ASCA 
(Grandi et al.\ 1997; Reynolds 1997; Sambruna et al.\ 1999),
(2) a relatively weak reflection component ($R$ = 0.65$\pm$0.12: 
see Table 4), determined with much better accuracy  
that observed with Beppo-SAX ($R$ = 0.56$^{+0.44}_{-0.20}$; 
Zdziarski \& Grandi 2001) and RXTE ($R$ = 0.4$^{+0.4}_{-0.1}$; 
Eracleous et al.\ 2000; see also Ballantyne et al.\ 2004 for updated
results) under the same constrained parameters, 
(3) the presence of a soft excess  
emission component below 3 keV, as firstly reported by Einstein 
(Turner et al.\ 1991) and by ROSAT (Grandi et al.\ 1997).
Finally, we confirmed the spectral variability of 3C~120 whereby the 
spectra becomes softer when it becomes brighter, as suggested by 
Maraschi et al.\ (1991) and Zdziarski \& Grandi (2001). 

The important question possibly raised by readers is  
$``what$ $is$ $completely$ $NEW$ $for$ $Suzaku?''$.  Firstly, 
we discovered, for the first time, the presence of a broad 
red-tail (asymmetric red wing) in the Fe K$_{\alpha}$ line of 
any radio galaxy. Interestingly, the line profile of 3C~120 is 
quite similar to those found in some Seyfert galaxies 
(e.g., MCG--5-23-16; Reeves et al.\ 2006), which provides important 
challenges to the unification models of radio-loud/quiet AGNs. 
We also detected a 6.9 keV line, 
but perhaps not as clearly as first noticed in the XMM-Newton data in 2003 
(Ballantyne et al.\ 2004). This may suggest significant changes 
in the Fe line profile on the year-long timescale, but further deep/long 
monitoring of 3C~120 is necessary for this kind of study.
Also, we have shown that Suzaku can provide a new 
diagnosis in accurate measurement of the line center 
energy ($\S$4.3), which was not possible before.

Second, we found the excess variance (i.e., variability amplitude) is 
generally larger at lower energies. From detailed multiband spectral 
studies, we  conclude the primary ``variable'' component is a steep 
($\Gamma$ = 2.6$-$2.7) power-law, which, at the same time, 
accounts for the soft excess emission below 3 keV. 
In fact, the difference spectrum is well represented 
by a steep power-law ($\Gamma_1$ = 2.65) modified by a flatter  
second component ($\Gamma_2$ = 1.75) at higher energies.
Apparently, the flat component mimics the ``direct'' power-law emission 
from the 3C~120 nucleus, and varies only a little (10 $\%$) during our 
observations. In contrast the flux of the steep power-law emission may 
have changed by a factor of $\ge$ 5  in the transition between the HF 
and LF states. Below we will discuss these new findings provided  
by Suzaku and their interpretation in detail.

\subsection{The Nature of the Iron K Line Complex}

Suzaku has successfully resolved the iron K line complex of 3C~120 
and also was first to verify the broad component's asymmetry. 
We showed that the iron line complex is composed of 
(1) a relatively narrow, neutral iron K line core,   
(2) broad iron line emission possibly emitted from 
the accretion disk (see $\S$ 4.3), and (3) an ionized $\sim$6.9 keV line. 
Interestingly, any of the models assuming a simple Gaussian profile 
(model-2, 4, 5 in Table 3) and/or an ``edge-on'' disk (model-5) 
predicts that the  centroid energy of the narrow line core is 
slightly shifted redwards by $\sim$ 30 eV. 
Meanwhile, the energy of the narrow core increases from 6.368 keV 
to 6.398 keV in the face-on model (model-6), which is then formally 
consistent with neutral iron at 6.40 keV. We suspect this is 
because the face-on diskline contributes towards some of the flux 
of the narrow 6.4 keV line core. For instance, the observed 30 eV 
shift of the line core implies a typical radius of 
$\sim$200 $r_g$ if the red-shift is purely gravitational. 

In contrast, the intrinsic line width of the narrow Fe line core was 
measured to be $\sigma$ $\simeq$ 110 eV. Assuming this line width is 
simply caused by Doppler broadening, this corresponds to a FWHM 
velocity of 10$^4$ km s$^{-1}$, indicating a possible origin in the 
broad line region (BLR). In the optical, the FWHM velocity of the BLR, 
as measured by Peterson et al.\ (2004) using the H$_{\beta}$ is 
smaller, 2200$\pm$200 km s$^{-1}$. Alternatively, Ogle et al.\ (2005) 
provides a very high S/N optical spectrum which shows a broader  
($\gg$ 2000 km s$^{-1}$) component associated with the BLR lines, which 
would be then more consistent with the width of the Fe line core if 
that also originates from the BLR. Thus we suggest most of the 
flux of the narrow 6.4 keV line core could be from the BLR, but that the
outer disk ($\sim$200 $r_g$) also makes some contribution.

The residuals present after subtracting the iron line core (Figure
\ref{fig:Fe-res} (b)) are poorly modeled either by adding a 
simple broad Gaussian or by adding the Compton shoulder of the 
K$_{\alpha}$ line, as we have discussed in $\S$4.3.
One may also suggest whether the presence of a warm absorber 
may affect continuum curvature to make an apparent red-tail below 
the iron K line.In this context, Ogle et al.\ (2005) present a very high S/N 
RGS spectrum, and there do not appear to be any significant lines or 
edges due to a warm absorber. So the warm absorber must be 
very weak or even absent in 3C~120 (but not surprising if we are 
viewing face-on). This also means that it is rather unlikely that any 
additional absorption can effect the Fe K line modeling, especially 
the red-wing (see also $\S$4.5). Rather, a significant red-tail below 
6.4 keV (in the rest frame of source) favors diskline 
emission from the inner accretion disk of $r_{\rm in}$ 
$\simeq$ 8.6$^{+1.0}_{-0.6}$ $r_g$. 
Indeed, adding the diskline emission provides a much better fit statistic 
than any other model as summarized in Table 3. If this broad line 
really originates from the inner accretion disk, it provides 
important clues to jet formation in the accretion disk.

For example, Reeves et al.\ (2006) have discovered a similar broad iron 
line in the Suzaku/XMM-Newton spectra  of MCG--5-23-16, which is thought to 
originate from inner accretion disk ($r_{\rm in}$ $\simeq$ 20 $r_g$). 
These observations may imply that both the radio-loud 3C~120 and the 
radio-quiet MCG--5-23-16 have similar accretion disk structure, 
in contrast to suggestions that the optically-thick accretion disk 
is truncated in 3C~120 to a hot, optically thin flow at a distance 
of  $r_{\rm in}$ $\sim$ 100 $r_g$ 
(Eracleous et al.\ 2000; Zdziarki \& Grandi 2001; Ballantyne et
al. 2004).  We also note that an EW of the diskline is 32$\pm$5 eV, 
whereas that for the narrow Fe line core is 60$\pm$6 eV.
Given the prediction of George and Fabian (1991) that 
a reflection parameter $R$ should be equal to EW(eV)/150(eV), 
the iron line EW is expected to be 90 eV for the level 
of Compton reflection $R$ $\simeq$ 0.6. Interestingly, 
this is consistent with the sum of the EWs of the narrow and diskline 
components, and therefore the Compoton hump could possibly be associated 
with both lines. 
Observations of the iron line profiles in various other broad line 
radio galaxies are important for the systematic comparison between  
Seyferts and BLRGs. We are planning to submit further deep observations
of other BLRGs in the next Suzaku observation program to test this.

Finally, to further test the robustness of the iron K diskline, 
we tried an alternative partial covering model, to see if that can 
reproduce the broad residuals present in the iron K band. 
Instead of the broad line, the neutral partial covering XSPEC model 
\textsc{pcfabs} was included, and the XIS+HXD data re-fitted over 
the range from 0.6$-$50 keV. The model assumes an absorbed double 
power-law continuum, together with a single Gaussian line
to model the iron K$\alpha$ line core, and ionized iron K emission as
described previously, while Compton reflection is also included in the
model with a cut-off energy of 100 keV. The partial covering model 
results in a fit statistic of 2999/2769 and leaves significant residuals 
around 6 keV in the iron K band. The fit statistic is statistically 
worse than the diskline fit ($\Delta\chi^{2}=21$ for 1 extra parameter), 
while the diskline has fit statistic of 2978/2768 (see $\S$4.5), 
and the equivalent width of the diskline is 45 eV. Without 
including either a diskline or a partial coverer, the fit statistic 
is 3018/2771. For a 50\% covering fraction, the upper-limit on 
the column density is then $<3\times10^{21}$\,cm$^{-2}$. Therefore 
the detection of the broad iron line appears to be robust in 3C\,120.

  \begin{figure}
  \begin{center}
    \FigureFile(85mm,85mm){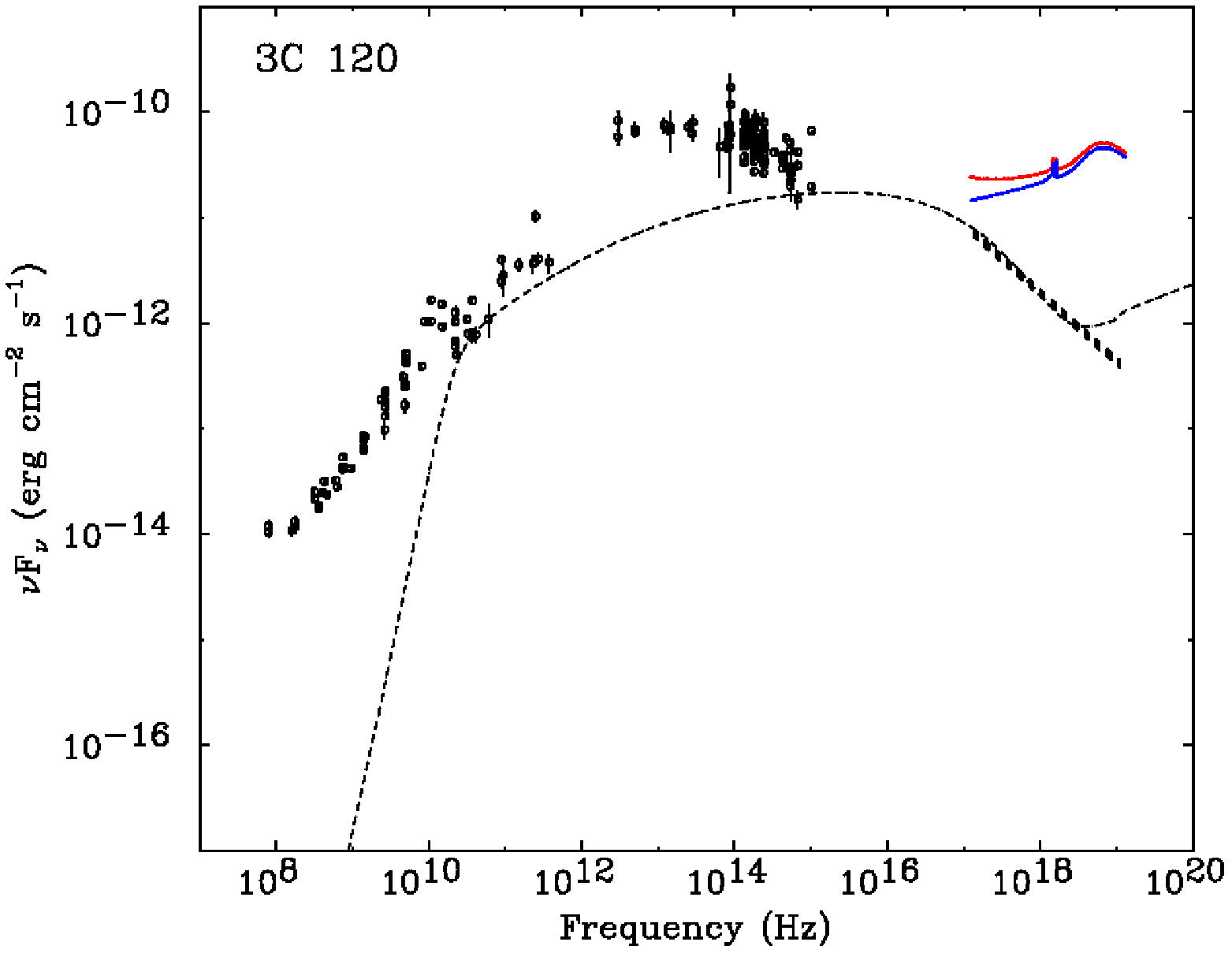}
    \FigureFile(85mm,85mm){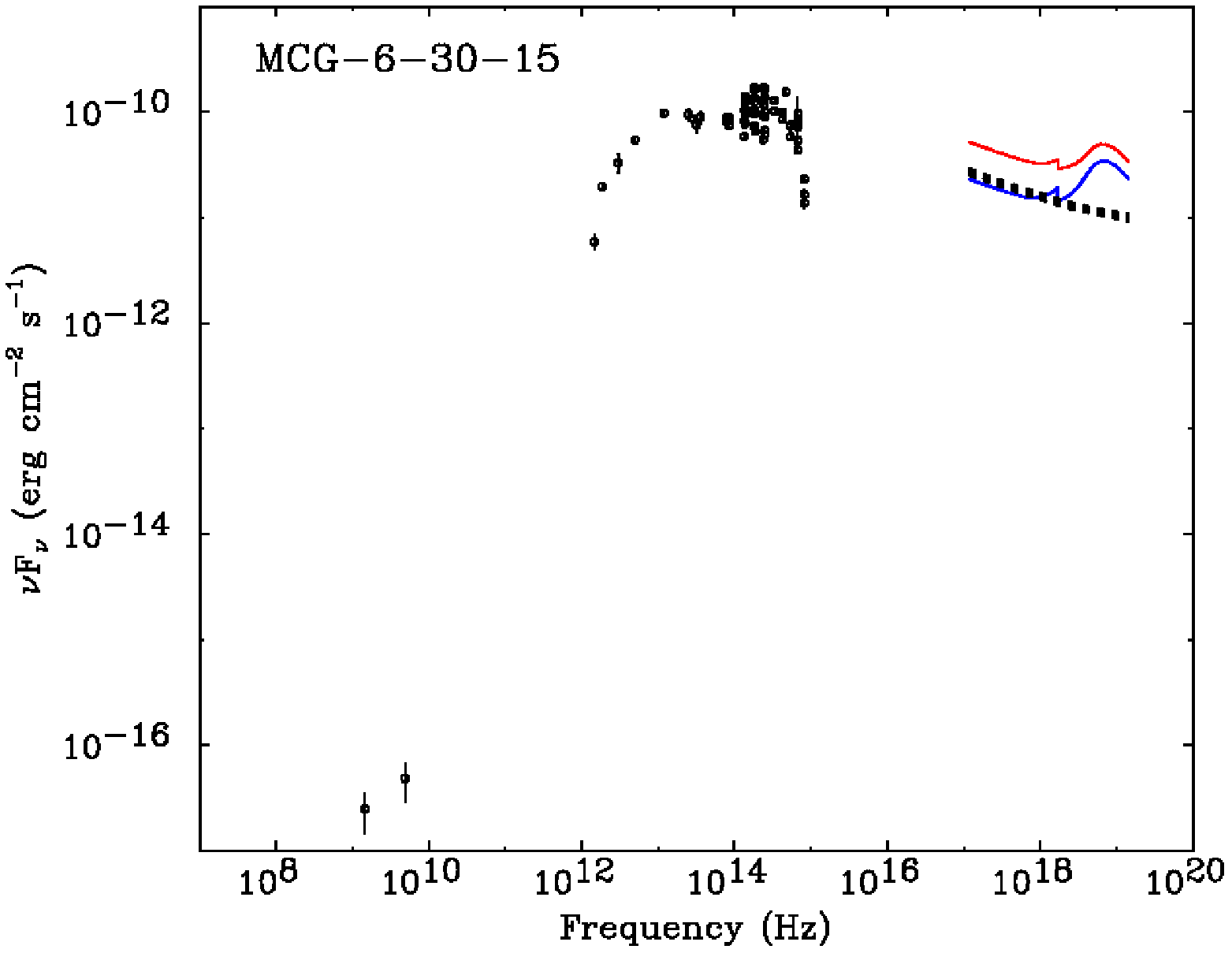}
  \end{center}
  \caption{Comparison of the multiband spectrum of a radio-loud 
  BLRG 3C~120 ($upper$) and a radio-quiet Sy-1 galaxy
  MCG--6-30-15 ($lower$). In both
   figures, X-ray data comes from the most recent Suzaku observations 
   (3C~120: this work. MCG--6-30-15: Minuitti et al.\ 2006) and the other 
   data comes from the NED data base. Red lines show the best-fit X-ray 
   spectral model during the high state, whereas blue lines shows that 
   for the low state. Variable power-law component is  shown as a thick dashed
   line for 3C~120, whereas the difference spectrum is shown for MCG--6-30-15. 
   The thin dotted line corresponds to an example fit of the 
    one-zone homogeneous SSC model as described in the text.}\label{fig:SEDcomp}
\end{figure}

\subsection {``Hidden'' Jet Emission? }

In $\S$ 4.5, we considered the origin of the spectral variability 
in 3C~120 by comparing the multiband spectra obtained in low flux 
(LF) and high flux (HF) states. Interestingly, several authors 
have examined similar spectral evolution of radio-quiet Seyfert
galaxies. For example, Miniutti et al.\ (2006) compared multiband spectra 
of the Sy-1 MCG--6-30-15  obtained in the LF and HF states. 
They found that difference spectrum shows a steep power-law of 
$\Gamma$ = 2.2, which is consistent with the direct power-law emission 
observed in both LF and HF states (see also Figure \ref{fig:SEDcomp}). 
This leads to an idea that the broad band spectral 
variability of MCG--6-30-15 is decomposed into two components; 
a highly variable power-law (direct nucleus emission) and a
constant reflection component plus iron line emission. Similarly, Reeves et al.\ (2006) 
presented a difference spectrum of a Sy-2 galaxy, MCG--5-23-16, 
which was fitted extremely well by a simple absorbed power-law of 
$\Gamma$ = 1.9. Again, this spectral shape is exactly consistent with 
the direct power-law component, which they observed in LF and 
HF states. Furthermore, there are no residuals present in the 
iron K band or any excess counts in the HXD/PIN difference spectrum 
above 10 keV.

In contrast to these findings for Seyfert galaxies, 
an interesting discovery made by Suzaku is that the variable component in 
3C~120 is much $steeper$ ($\Gamma$ $\simeq$ 2.7) than the power-law 
emission reported in literature (1.6 $\le$ $\Gamma$ $\le$ 1.8: 
see also Marscher et al.\ 2002).  
One interesting idea to account for this steep, variable emission 
is the beamed radiation from the jet (not on kpc scales as discussed 
in Harris et al.\ 2004 and $\S$1, but the unresolved base of the jet 
on sub-pc scales), though this component contributes only 
$\sim$ 20$\%$ at most, of the Sy-1 like X-ray emission 
in 3C~120 (i.e., emitted from the disk and corona). In fact, 3C~120 
possesses a superluminal radio jet with an inclination angle $\le$ 
14 deg (Eracleous \& Halpern 1988). This low inclination angle implies 
that 3C~120 may have some ``blazar-like'' characteristics, such as 
rapid X-ray variability or a non-thermal spectrum extending to the 
$\gamma$-ray energy band. 
For example, if we assume a jet bulk Lorentz factor of 
$\Gamma_{\rm BLK}$ = 10 and a jet inclination angle of 14 deg, 
we expect the observed flux is mildly ``boosted'' towards the observer with a 
Doppler beaming factor $\delta$ $\sim$ 3. Note also that the X-ray 
spectra of blazars are generally represented by a power-law function, 
but their X-ray photon indices ranges widely (1.5 $\le$ $\Gamma$ $\le$ 3.0; 
Kubo etal. 1998) among various sub-classes 
(e.g., HBL, LBL, and FSRQ; e.g., Fossati et al.\ 1998).
Interestingly, similarly variable, steep power-law emission 
has been found for a distant ($z$ = 0.94) quasar PG1407+265 which is 
classified as a radio-quiet object, but shows strong evidence 
of a radio jet with a highly relativistic speed in the 
VLBA observations (Gallo 2006).

Figure \ref{fig:SEDcomp} shows the multi-band spectrum of 3C~120 ($upper$), 
compared to that of the radio-quiet Sy-1 galaxy MCG--6-30-15 ($lower$) 
between the radio and X-ray energy bands. Note the significant
difference in the radio band 
(as inferred by a definition of radio-quiet/loud objects), 
but spectral similarity in the optical band. 
In both figures, the X-ray data comes from the recent 
Suzaku observations. Red/blue lines show the best-fit 
X-ray spectral model during HF/LF states. The difference spectrum  
is shown as a thick dashed line. The origin of the variable
component (steep power-law) in 3C~120 is still uncertain, 
but here we assume the non-thermal jet dominates this power-law
emission. The thin dotted line shows an example fit with a 
one-zone homogeneous SSC (Synchrotron self-Compton) model 
described in Kataoka et al.\ (1999), where the soft X-ray 
excess corresponds to the highest energy end of the synchrotron
emission. 

Also above 10 keV, the inverse Compton component 
is possibly present even in the hard X-ray band, but the isotropic 
Seyfert-like emission overwelms that blazar-like emission, 
where the relativistic boost of the jet is smaller than in blazars, 
because of a lower $\delta$. Note that this model for the 
jet-like emission - with the SED peaking at $\sim$ 
3 $\times$ 10$^{15}$ $-$ 10$^{16}$ Hz - 
implies HBL-like classification of the jetted emission.  
This in principle fits well with the FR-I type classification of 
the object discussed in Section 1, where HBL blazars are FR-I 
radio galaxies viewed close to the direction of the jet.  However, there is
an additional complication:  the FR-I / HBL nature of this object as 
inferred from the radio observations and the soft X-ray excess is 
in conflict with the presence of broad emission lines, present in the 
optical spectrum, as those are generally {\it not} observed in 
FR-I sources.

Interestingly, the jet parameters derived here are consistent with 
typical values for blazars, except for a moderate beaming factor of  
$\delta$ = 3 (viz  5 $\le$ $\delta$ $\le$ 30 for blazars; e.g., 
Kubo et al.\ 1998): magnetic field $B$ = 0.3 G, 
region size $R$ = $c t_{\rm var}$ $\delta$ $\sim$ 
1.0$\times$10$^{17}$cm, under an assumption of equipartition between 
electron and field energy densities ($u_{\rm e}$ = $u_{\rm B}$). 
We assumed a broken power-law form of the 
electron distribution $N(\gamma)$ $\propto$ $\gamma^{-s}$
$(1+\gamma/\gamma_{\rm br})^{-1}$ exp($- \gamma$/$\gamma_{\rm max}$), 
where $\gamma$ is the electron energy (in units of $m_e c^2$), 
$\gamma_{\rm brk}$ = 10$^3$, $\gamma_{\rm max}$ 
= 1.3$\times$10$^5$, and $s$ = 1.8.  Surprisingly, 
this simple jet model reproduces the general trend of the spectral 
energy distribution quite well, though some discrepancies are seen 
in the radio band. Such discrepancies 
in the radio are found in many studies and it is believed to be a 
consequence of the radio emission originating from a much larger 
region than the X-rays (e.g., Kataoka et al.\ 1999).  
Also note that the prediction of gamma-ray flux via  
inverse Comptonization is less than 10$^{-10}$ 
erg cm$^{-2}$ s$^{-1}$, qualitatively consistent with the
$non$-detection by EGRET and COMPTEL onboard CGRO.   

Based on these facts alone, it is still premature to conclude that the 
steep variable X-ray emission is actually originating from the jet in
3C~120. Maraschi et al.\ (1991) and  Zdziarski \& Grandi (2001) claimed
that an X-ray jet component could, at the very least, dilute the 
Seyfert-like spectrum of 3C~120, and may account for the observed weak 
reflection features. This idea is straightforward, but may be 
oversimplified  by following reasons: 
(1) the Suzaku observations have now revealed that the variable emission can
explain at most 20$\%$ of the Sy-1 emission even below 3 keV. 
(2) The putative jet spectrum is much steeper than the direct power-law 
($\Gamma$ $\simeq$ 1.7). Such a low amplitude, steep 
power-law jet component cannot dilute the Sy-1 emission sufficiently,
especially above 5 keV (where the iron lines and reflection 
become important, unless there is additional, weak contribution from a 
hard spectrum due to the inverse Compton component.
\footnote{In this context, we note that an inverse Compton model 
in Figure \ref{fig:SEDcomp} is dependent on input parameters 
as $f_{\rm IC}$ $\propto$ $u_{\rm B}^{-2}$, where $f_{\rm IC}$ is an 
observed inverse Compton flux. Therefore this could dilute the 
Sy-1 emission only if the magnetic field strength is about a factor of 
five $smaller$ than equipartition values, which is often not the 
case for blazars (see, e.g., Kubo et al. 1998).}
(3) No evidence for ``blazar-like'' variability has been found  
above 2 keV (Giozzi et al.\ 2003; Marshall et al.\ 2003). Therefore, 
it seems that the weak iron line and the reflection hump  
observed in BLRGs (Eracleous et al.\ 2000) might be more intrinsic, 
and provides important clues to the origin of the jet-like emission.

Nevertheless, we showed that the variable steep X-ray emission is a 
key to understanding the spectral evolution of 3C~120. In particular, it
seems that the steep power-law component is more significant when the source 
is brighter, and can well explain the significant soft excess repeatedly 
observed in the literature for this AGN. Interestingly, the soft excess 
emission is hardly visible when the source becomes fainter, which may 
indicate that 
the jet component has completely disappeared and/or is hidden behind much 
stronger Sy-1 emission. Unfortunately, the flux changes of 3C~120 during 
the Suzaku observation were relatively small, so we cannot conclude what 
fraction of X-ray flux is actually explained by the jet.  
Future deep Suzaku observations,  as well as continuing VLBI monitoring 
coincident with X-ray monitoring (as the campaigns reported 
by Marscher et al. 2002), sensitive measurements with GLAST, 
in quite different states of source activity (i.e., observations 
at  historically high and low states) will be crucial to 
understanding the nature of emission properties in 3C~120 
(see recent paper by Grandi and Palumbo (2006) for the 
detectability with GLAST at MeV$-$GeV energy band).

\section{Conclusion}

We have presented a detailed analysis of results from the broad line radio 
galaxy 3C~120 observed with Suzaku from February and March 2006. 
Thanks to the excellent sensitivity 
of both the XIS and HXD/PIN detectors onboard Suzaku, we obtained multi-band data with
unprecedented accuracy between 0.6 and 50 keV. Our major findings are as 
follows:\\
(1) An overall spectral shape, including weak reflection 
($R$ $\simeq$ 0.6) and a soft excess which was consistent with previous findings, 
and was determined with unprecedented accuracy.\\ 
(2) We confirmed the presence of a narrow iron K$_{\alpha}$ core 
with a width of $\sigma$ = 111$^{+11}_{-10}$  eV (and an EW = 60$\pm$6
eV), and 
a 6.93$\pm$0.02 keV emission line possibly emitted from H-like iron. \\
(3) After subtracting these narrow line components, significant residuals
remain redwards of the narrow iron line core, well below 6.4 keV.\\
(4) The difference spectrum between LF and HF clearly shows a power-law 
of $\Gamma$ = 2.2, which is naturally interpreted as a combination 
of a highly variable steep power-law ($\Gamma$ =~ 2.7; jet-like) plus a
moderately variable direct power-law emission ($\Gamma$ =~ 1.7; Sy-1
like). No significant variability was found in either the Fe K line
emission or reflection component.\\

We argue that the narrow iron K line is primarily emitted from the broad 
line region corresponding to a FWHM velocity of 10$^4$ km/s, 
but the emission from the outer disk ($\sim$ 200 $r_g$) may also 
play a part. Meanwhile, the broad redshifted iron line
can be interpreted as emission 
from the inner, face-on accretion disk  ($r_{\rm in}$ $\sim$ 10$r_g$ and 
$i$ $\le$ 10 deg). We considered the origin of a highly variable 
steep power-law component in the context of a possible 
relation to non-thermal jet emission. 
Although we cannot conclude what fraction of X-ray
emission in 3C~120 is actually explained by the jet, it must be 
less than 20 $\%$ of the ``Sy-1 like'' emission even below 3 keV. 
This clear difference of spectral evolution between Sy-1s and 3C~120
provides an important motivation for further future deep observations 
of BLRGs. 

Finally, we thank an anonymous referee for his/her constructing 
comments which helped clarify many of the issues presented in this
paper. We also thank Dr. Tahir Yaqoob  for his helpful comments and 
discussion on the Suzaku data analysis.

\appendix
\section{Effect of Background Variations on the HXD/PIN Light Curve}

\begin{figure}
  \begin{center}
    \FigureFile(68mm,80mm){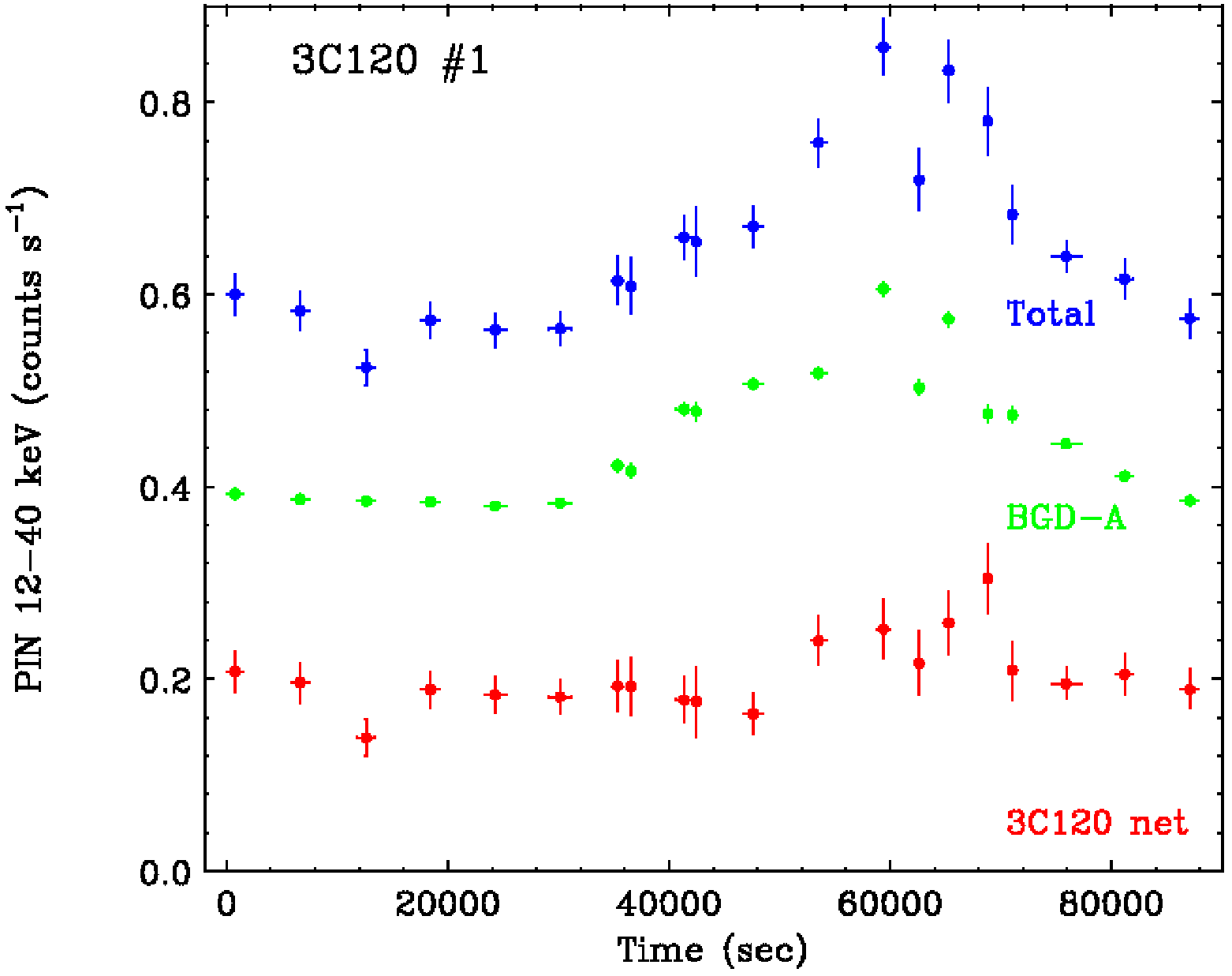}
    \FigureFile(68mm,80mm){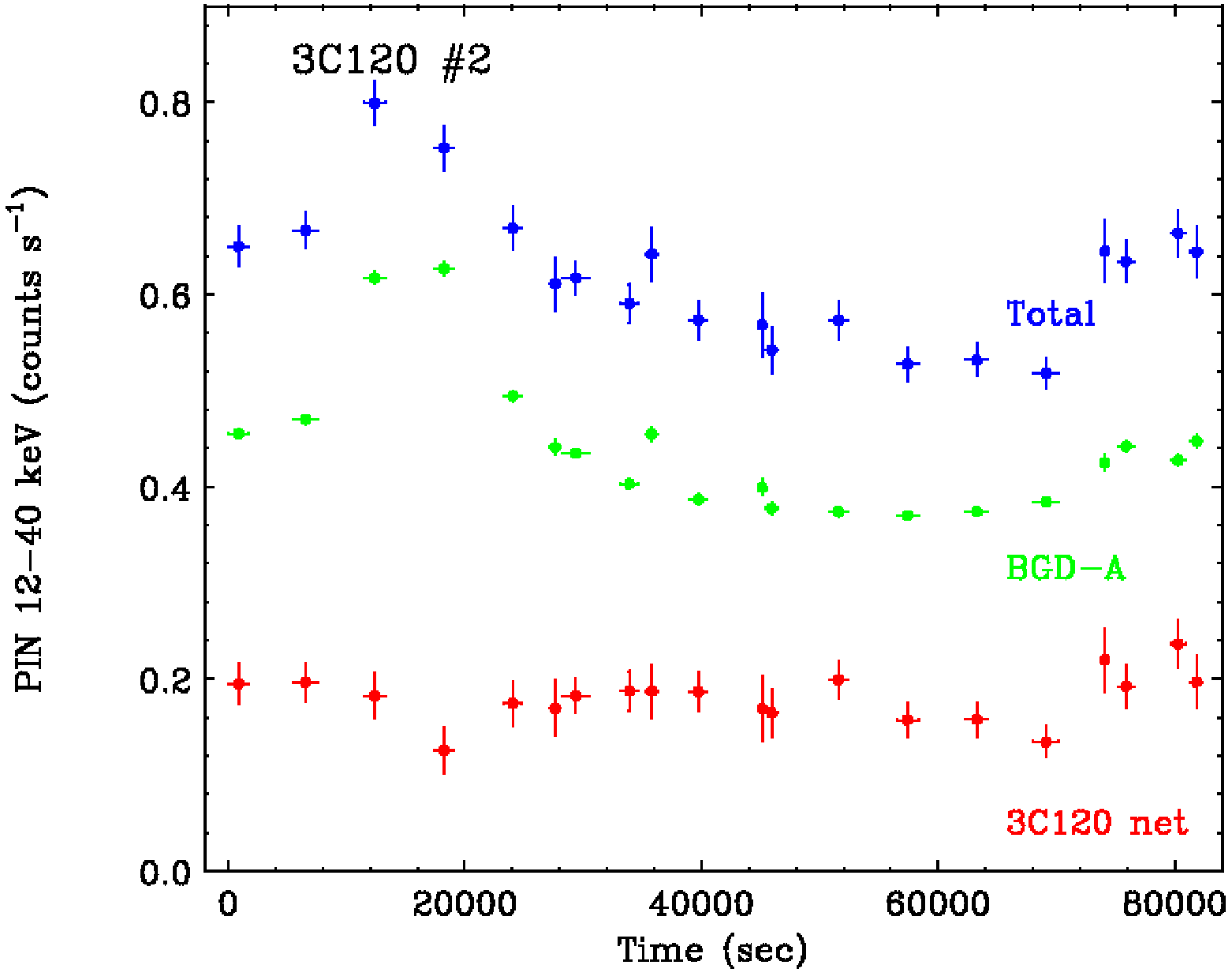}
    \FigureFile(68mm,80mm){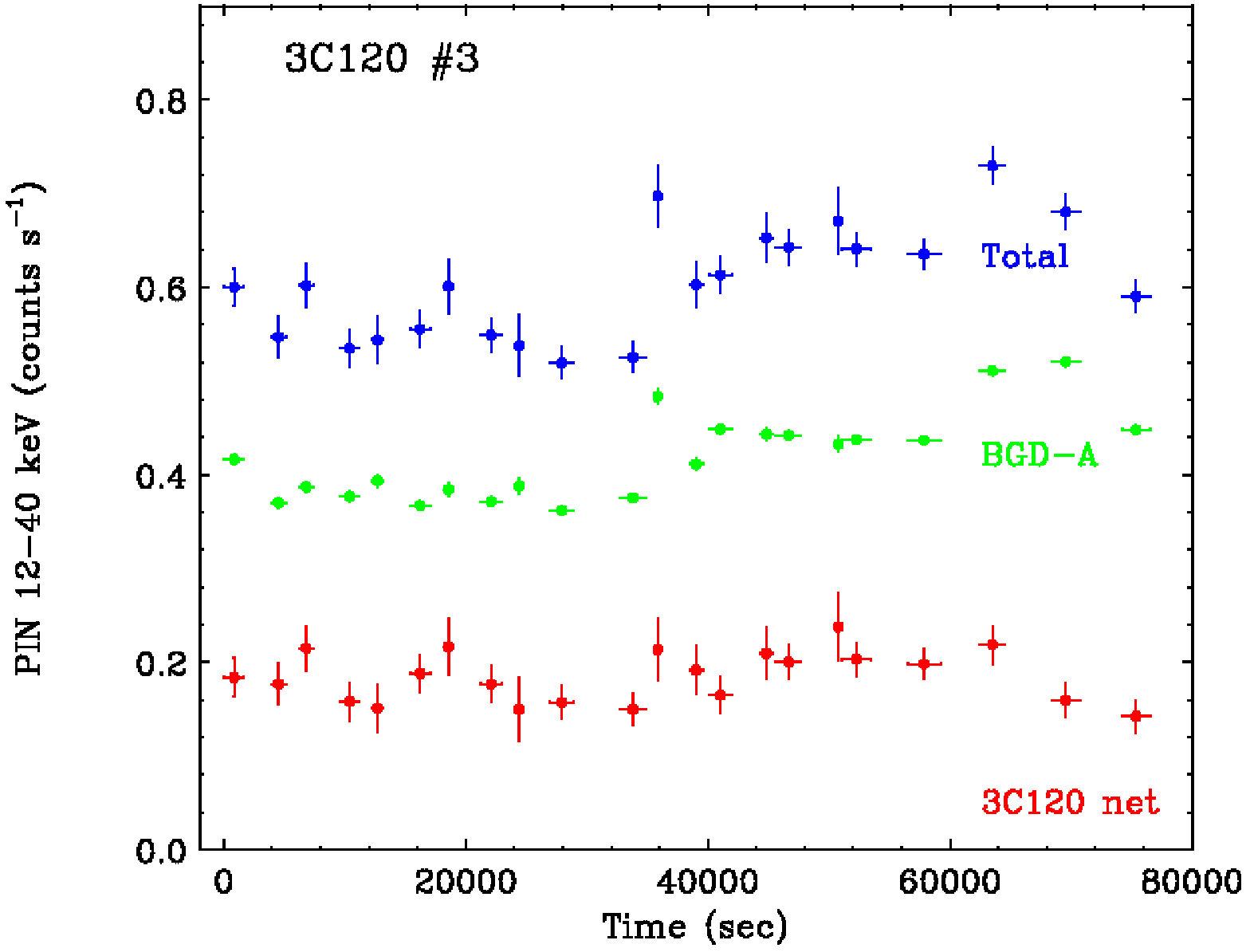}
    \FigureFile(68mm,80mm){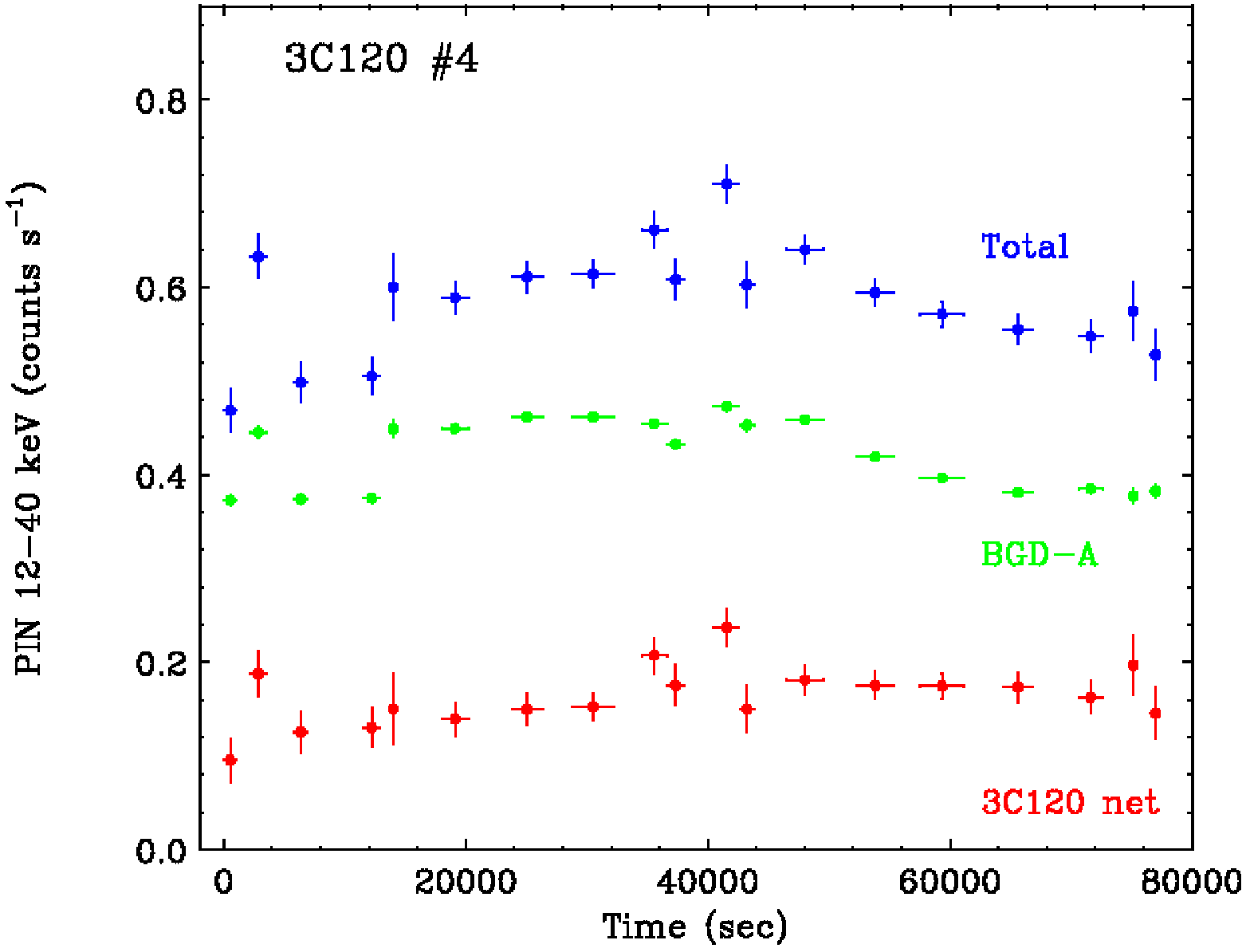}
  \end{center}
  \caption{Comparison of temporal variation of total 
HXD/PIN counts ($blue$: including source and background), 
non X-ray background model A ($green$; see $\S$2.2), and resultant  
estimate of source counts from 3C~120 ($red$) for 3C~120 $\#1 - 4$.}
\label{fig:pin-bgd-lc}
\end{figure}

Here we provide a follow-up discussion concerning the robustness 
of background subtraction of the HXD/PIN using the most recent 
response and background model as of November 2006 (v1.2 data 
and ae\_hxd\_pinxinom\_20060814.rsp).
Detailed studies of background systematics are 
still under investigation by the HXD instrumental team, but  
careful comparison of the HXD/PIN light curves (see Figure \ref{fig:lc}) 
with background variations may provide important hints for future 
modeling of relatively faint sources. For this aim, 3C~120 is a
good target as it is bright enough to be firmly detected by the 
HXD/PIN, but much fainter than the background photon statistic. 
As we have seen in Figure \ref{fig:pinspec}, the net intensity of 3C~120 
is expected to be $\sim$30 $\%$ of the non X-ray background plus CXB above 
12 keV.  

Figure \ref{fig:pin-bgd-lc} compares temporal variations of total 
HXD/PIN counts ($blue$: including source and background), 
non X-ray background model A ($green$; see $\S$2.2), and resultant  
estimate of source counts from 3C~120 ($red$).  The HXD/PIN count 
rate ($blue$) is highly variable from 0.5 counts s$^{-1}$ to 
0.9 counts s$^{-1}$, and this trend is generally well reproduced 
in the background model ($green$). The dead time is corrected for 
each of the time bins, because it varies with 
the event rate (dominated by background photons) of the HXD/PIN
detector and ranges from 3.5$\%$ to 8.5$\%$ during 3C~120 $\#1 - 4$ 
observations. As a result, small residuals (of 5$-$10 $\%$ background
level)  remains after subtracting the background photons, especially 
when the total HXD/PIN counts exceeds $\sim$ 0.7 counts s$^{-1}$. 
Since these time regions correspond to orbits where the 
COR is low and/or passing through the SAA, a long-decay component of 
the background may have not been modeled perfectly in the current 
data. 
 
Similarly, the relative lack of variability in the XIS light curve 
suggests apparent ``rapid'' changes in HXD/PIN flux   
Figure \ref{fig:pin-bgd-lc} could be artifacts of the background 
subtraction.
Nevertheless, the general trend of the HXD/PIN flux variations 
traces well those of the XIS data below 10 keV, as 
we have seen in Figure \ref{fig:hd-pin}. The average net source count 
rate of the HXD/PIN is 0.195$\pm$0.005 
cts s$^{-1}$, 0.177$\pm$0.005 cts s$^{-1}$, 0.180 $\pm$0.004 cts s$^{-1}$, 
and 0.165$\pm$0.004 cts s$^{-1}$ respectively, and hence the 
maximum is in obs $\#$1 and the minimum in  $\#$4, as per the XIS. 
A constant fit to overall HXD/PIN light curves (obs $\#1 - 4$) provides 
$\chi^2$/d.o.f = 154/80. Even after removing the time regions 
where the background is high (HXD/PIN counts exceeds 
0.7 counts s$^{-1}$), 
we obtain $\chi^2$/d.o.f = 103/71 ($P(\chi^2)$ = 0.7 $\%$).
Therefore we think the long-term variations of 3C~120 are robust in 
the 12$-$40 keV energy band. More detailed discussion using revised 
background models will be discussed elsewhere.


\end{document}